\documentclass[twoside,twocolumn,9pt]{article}
\usepackage{extsizes}
\usepackage[super,sort&compress,comma]{natbib} 
\usepackage[version=3]{mhchem}
\usepackage[left=1.5cm, right=1.5cm, top=1.785cm, bottom=2.0cm]{geometry}
\usepackage{balance}
\usepackage{mathptmx}
\usepackage{sectsty}
\usepackage{graphicx} 
\usepackage{lastpage}
\usepackage[format=plain,justification=justified,singlelinecheck=false,font={stretch=1.125,small,sf},labelfont=bf,labelsep=space]{caption}
\usepackage{float}
\usepackage{fancyhdr}
\usepackage{fnpos}
\usepackage[english]{babel}
\addto{\captionsenglish}{%
  \renewcommand{\refname}{Notes and references}
}
\usepackage{color,soul}
\usepackage{multicol}
\usepackage{comment}
\usepackage{array}
\usepackage{droidsans}
\usepackage{charter}
\usepackage[T1]{fontenc}
\usepackage[usenames,dvipsnames]{xcolor}
\usepackage{setspace}
\usepackage[compact]{titlesec}
\usepackage{hyperref}
\usepackage{booktabs}
\usepackage{multirow}
\usepackage{amssymb}
\usepackage{placeins}
\newcommand{\beginsupplement}{%
    \setcounter{figure}{0}%
    \renewcommand{\thefigure}{S\arabic{figure}}%
    \setcounter{table}{0}%
    \renewcommand{\thetable}{S\arabic{table}}%
}

\usepackage{epstopdf}

\definecolor{cream}{RGB}{222,217,201}

\begin{document}

\pagestyle{fancy}
\thispagestyle{plain}
\fancypagestyle{plain}{
\renewcommand{\headrulewidth}{0pt}
}

\makeFNbottom
\makeatletter
\renewcommand\LARGE{\@setfontsize\LARGE{15pt}{17}}
\renewcommand\Large{\@setfontsize\Large{12pt}{14}}
\renewcommand\large{\@setfontsize\large{10pt}{12}}
\renewcommand\footnotesize{\@setfontsize\footnotesize{7pt}{10}}
\makeatother

\renewcommand{\thefootnote}{\fnsymbol{footnote}}
\renewcommand\footnoterule{\vspace*{1pt}%
\color{cream}\hrule width 3.5in height 0.4pt \color{black}\vspace*{5pt}} 
\setcounter{secnumdepth}{5}

\makeatletter 
\renewcommand\@biblabel[1]{#1}            
\renewcommand\@makefntext[1]%
{\noindent\makebox[0pt][r]{\@thefnmark\,}#1}
\makeatother 
\renewcommand{\figurename}{\small{Fig.}~}
\sectionfont{\sffamily\Large}
\subsectionfont{\normalsize}
\subsubsectionfont{\bf}
\setstretch{1.125} 
\setlength{\skip\footins}{0.8cm}
\setlength{\footnotesep}{0.25cm}
\setlength{\jot}{10pt}
\titlespacing*{\section}{0pt}{4pt}{4pt}
\titlespacing*{\subsection}{0pt}{15pt}{1pt}

\fancyfoot{}
\fancyfoot[LO,RE]{\vspace{-7.1pt}\includegraphics[height=9pt]{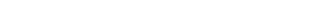}}
\fancyfoot[CO]{\vspace{-7.1pt}\hspace{13.2cm}\includegraphics{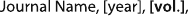}}
\fancyfoot[CE]{\vspace{-7.2pt}\hspace{-14.2cm}\includegraphics{head_foot/RF}}
\fancyfoot[RO]{\footnotesize{\sffamily{1--\pageref{LastPage} ~\textbar  \hspace{2pt}\thepage}}}
\fancyfoot[LE]{\footnotesize{\sffamily{\thepage~\textbar\hspace{3.45cm} 1--\pageref{LastPage}}}}
\fancyhead{}
\renewcommand{\headrulewidth}{0pt} 
\renewcommand{\footrulewidth}{0pt}
\setlength{\arrayrulewidth}{1pt}
\setlength{\columnsep}{6.5mm}
\setlength\bibsep{1pt}

\makeatletter 
\newlength{\figrulesep} 
\setlength{\figrulesep}{0.5\textfloatsep} 

\newcommand{\topfigrule}{\vspace*{-1pt}%
\noindent{\color{cream}\rule[-\figrulesep]{\columnwidth}{1.5pt}} }

\newcommand{\botfigrule}{\vspace*{-2pt}%
\noindent{\color{cream}\rule[\figrulesep]{\columnwidth}{1.5pt}} }

\newcommand{\dblfigrule}{\vspace*{-1pt}%
\noindent{\color{cream}\rule[-\figrulesep]{\textwidth}{1.5pt}} }

\makeatother

\twocolumn[
  \begin{@twocolumnfalse}
{\includegraphics[height=30pt]{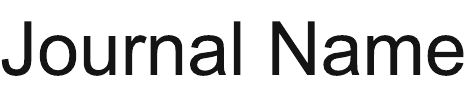}\hfill\raisebox{0pt}[0pt][0pt]{\includegraphics[height=55pt]{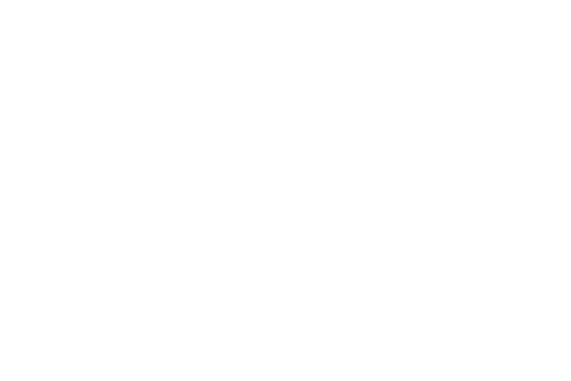}}\\[1ex]
\includegraphics[width=18.5cm]{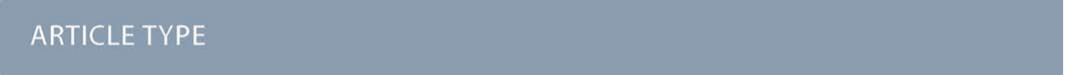}}\par
\vspace{1em}
\sffamily
\begin{tabular}{m{4.5cm} p{13.5cm} }

\includegraphics{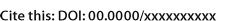} & \noindent\LARGE{\textbf{Electronic manifolds for extrapolative alloy discovery}} \\
\vspace{-0.3cm} & \vspace{-0.3cm} \\

  & \noindent\large{Pranoy Ray\textit{$^{a, b}$}, Sayan Bhowmik\textit{$^{c}$}, Phanish Suryanarayana\textit{$^{b, d}$}, Surya R. Kalidindi\textit{$^{a, b}$} and Andrew J. Medford\textit{$^{c\dag}$}} \\

\includegraphics{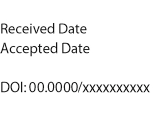} & \noindent\normalsize{This study presents a computationally efficient framework for accelerated alloy discovery that uses the non-interacting electron density to capture intrinsic structure-property relationships in refractory high-entropy alloys (HEAs). Unlike state-of-the-art approaches relying on expensive, self-consistent density functional theory calculations, our method employs the non-interacting electron density as the primary structural descriptor. By extracting physical features through directionally resolved two-point spatial correlations and compressing them via Principal Component Analysis, we efficiently map the design space. Coupling these descriptors with Bayesian active learning, we achieve a normalized mean absolute error (NMAE) of <2\% for the bulk modulus of Al-Nb-Ti-Zr alloys using only 10 training samples (<0.2\% of the dataset). Furthermore, we demonstrate that the model learns an electronic packing manifold that is transferable within the refractory BCC alloy family. Validated on a distinct 7-component refractory system (Mo-Nb-Ta-Ti-V-W-Zr) containing four elements entirely absent from the training data, the framework enables zero-shot transfer within the refractory BCC alloy class. Moreover, by augmenting the base model with just 20 samples from the target domain, we achieve high-fidelity predictions (NMAE $<$ 3\%) for 7-component alloys, reducing data acquisition costs by orders of magnitude compared to standard workflows. A controlled comparison confirms that composition-based descriptors under the identical pipeline do not reach the same accuracy threshold within the same sample budget, establishing that the spatial autocorrelation encoding of the non-interacting electron density provides information beyond elemental composition statistics alone.} \\

\end{tabular}

 \end{@twocolumnfalse} \vspace{0.6cm}

  ]

\renewcommand*\rmdefault{bch}\normalfont\upshape
\rmfamily
\section*{}
\vspace{-1cm}

\begin{figure*}[h!]
\includegraphics[width=\textwidth]{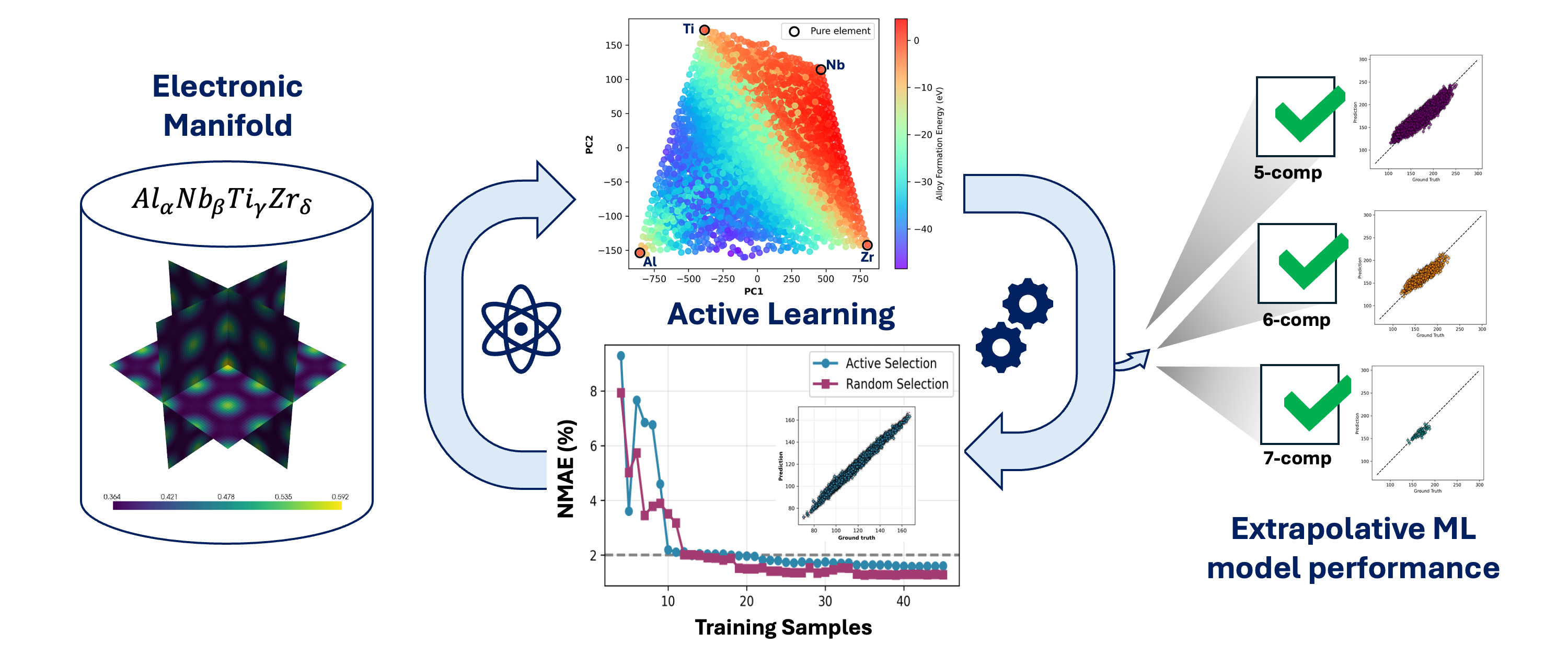}
\caption*{Graphical Abstract}
\end{figure*}

\footnotetext{\textit{$^{a}$~George W. Woodruff School of Mechanical Engineering, Georgia Institute of Technology, Atlanta, GA, USA}}
\footnotetext{\textit{$^{b}$~School of Computational Science and Engineering, Georgia Institute of Technology, Atlanta, GA, USA}}
\footnotetext{\textit{$^{c}$~School of Chemical and Biomolecular Engineering, Georgia Institute of Technology, Atlanta, GA, USA}}
\footnotetext{\textit{$^{d}$~School of Civil and Environmental Engineering, Georgia Institute of Technology, Atlanta, GA, USA}}
\footnotetext{\dag~Corresponding Author: ajm@gatech.edu}





\section*{Introduction}

High-entropy alloys (HEAs)\cite{jeje_emerging_2025, dang_elastocaloric_2025, tabor_accelerating_2018} encompass a vast compositional design space that offers exceptional tunability for mechanical and thermal properties, yet this combinatorial magnitude renders exhaustive experimental exploration intractable. While Kohn-Sham density functional theory (DFT)\cite{kohn_self-consistent_1965, hohenberg_inhomogeneous_1964} offers first-principles accuracy\cite{chakraborty_high_2021, kundu_zr_2024, nair_ti-decorated_2023}, its large computational cost and $\mathcal{O}(N^3)$ scaling with system size render exhaustive high-throughput screening\cite{ramakrishna_materials_2019, ray_lean_2025} of HEA compositional spaces computationally prohibitive. Machine learning (ML) surrogates\cite{ward_general-purpose_2016, ward_matminer_2018, ray_ml_2026, lei_universal_2022, kaundinya_prediction_2022, kalidindi_vision_2016, brockherde_bypassing_2017, xie_crystal_2018, choudhary_atomistic_2021, merchant_scaling_2023, bartok_representing_2013} address this by approximating property predictions at reduced cost. Recent efforts have further enhanced sample efficiency through Bayesian active learning\cite{lyngby_bayesian_2024, kusne_--fly_2020, qian_knowledge-driven_2023, lookman_active_2019, ray_refining_2025, ray_assessing_2025, buzzy_active_2025, khatamsaz_bayesian_2023, nakayama_tuning_2022, startt_bayesian_2024, hou_bayesian_2020, frazier_tutorial_2018, alvi_hierarchical_2025} and Gaussian Process Regression (GPR)\cite{hastie_elements_2001, rasmussen_gaussian_2005, boender_bayesian_1991, mockus_bayesian_1989, hanaoka_bayesian_2021, seyed_mahmoud_sequential_2026}, though their efficacy depends critically on the choice of structural descriptors.      

Recent advances\cite{zhao_predicting_2020, casey_prediction_2020, barry_voxelized_2023-1, ray_lean_2025} in physics-based feature engineering have established the electron density field as a robust, chemically agnostic descriptor for atomic structures. These methods are powerful because the converged charge density encodes the quantum mechanical ground state, but this fidelity comes at substantial computational cost. State-of-the-art frameworks, such as the Voxelized Atomic Structure (VASt) method\cite{barry_voxelized_2020, barry_voxelized_2023, qu_voxelized_2023}, utilize the fully converged elecron density field to quantify structural features via two-point spatial correlations\cite{kalidindi_hierarchical_2015, kaundinya_machine_2021, cecen_material_2018, ray_lean_2025, ray_ml_2026, mann_development_2022}. While these descriptors achieve high fidelity, they create a fundamental inefficiency: the calculation of the electronic ground state via self-consistent field (SCF) iteration, the dominant computational cost in static DFT, must be completed for every candidate structure merely to generate input features for the surrogate model. This limits the primary value proposition of machine learning surrogates, as the computational budget required for feature generation becomes comparable to that of directly computing the target property.

To address this bottleneck, we investigate the efficacy of the non-interacting valence electron density, hereafter referred to as pseudo-density. In practice, the pseudo-density represents the superposition of isolated valence electron densities corresponding to atoms placed at the Special Quasirandom\cite{zunger_special_1990} (SQS) lattice sites, determined via Vegard's Law to bypass equilibrium geometry optimization, with no subsequent electronic relaxation. By eliminating the iterative SCF cycle, the presented approach reduces the computational cost of feature generation by orders of magnitude compared to fully converged DFT\cite{barry_voxelized_2023}, while preserving essential chemical and valence electron information. Our approach rests on the premise that in HEAs, property variance is driven by chemical composition wherein superimposed valence densities could act as an effective descriptor. Consequently, the pseudo-density should retain sufficient physical fidelity for predictive modeling; a hypothesis supported by recent work\cite{lei_universal_2022}. Furthermore, because these descriptors are derived via unsupervised learning, solely from atomic configuration and elemental identity independent of any target property, they form a property-agnostic representation capable of predicting disparate physical quantities, e.g., mechanical and thermodynamic properties, without feature re-engineering. Crucially, we posit that this physics-based encoding offers a route to transferability: because pseudo-electron densities capture the spatial packing and overlap of valence electrons, they may encode a transferable electronic manifold within structurally and chemically constrained alloy families. This would enable a model trained on a lower-order system, e.g., 4-component, to achieve predictive transfer to higher-order systems, e.g., 7-component, within the same alloy class with minimal additional data.

In this work, we present a framework integrating pseudo-density descriptors: quantified via two-point spatial correlations and PCA, with Gaussian Process Regression (GPR) driven Bayesian active learning. We validate the approach on the Al-Nb-Ti-Zr system ($\mathcal{D}_4$), achieving $R^2 > 0.97$ for both bulk modulus and alloy formation energy. Notably, the bulk modulus model attains a normalized mean absolute percentage error (NMAE) of $<2\%$ using only 10 training samples, surpassing benchmarks utilizing converged densities. We further demonstrate extrapolative power by applying the $\mathcal{D}_4$-trained model to a distinct 7-component system ($\mathcal{D}_7$: Mo-Nb-Ta-Ti-V-W-Zr). The framework enables zero-shot transfer within the refractory BCC alloy class, and with the augmentation of just 20 actively selected $\mathcal{D}_7$ samples, recovers high-fidelity predictions (NMAE $< 3\%$), establishing pseudo-density as a practical descriptor for sample-efficient refractory alloy discovery.

\section*{Data Curation and Design Space}
This study investigates the quaternary Al-Nb-Ti-Zr refractory HEA system, spanning a large compositional space defined by:
\begin{align}
\mathcal{D}_4
= \bigl\{\mathrm{Al}_{\alpha}\mathrm{Nb}_{\beta}\mathrm{Ti}_{\gamma}\mathrm{Zr}_{\delta}
\mid &\alpha,\beta,\gamma,\delta \in \{0,4,\dots,128\}, \bigr. \notag \\
&\bigl. \alpha+\beta+\gamma+\delta = 128 \bigr\}.
\label{eq:composition}
\end{align}
This discrete set yields a total of 6,545 unique structures when including elemental, binary, ternary, and quaternary compositions, and 4,495 alloys when restricted to the quaternary space. To validate the extrapolative transferability of the model, we leverage the distinct 7-component domain ($\mathcal{D}_7$) which contains 12012 unique compositions, defined by the set:
\begin{align}
\mathcal{D}_7
= & \bigl\{\mathrm{Mo}_{\alpha}\mathrm{Nb}_{\beta}\mathrm{Ta}_{\gamma}
\mathrm{Ti}_{\delta}\mathrm{V}_{\epsilon}\mathrm{W}_{\zeta}\mathrm{Zr}_{\eta}
\mid  \notag \\
& \alpha,\dots,\eta \in \{0,15,23,30,38,\dots,105,128\}, \bigr. \notag \\
&\bigl. \alpha+\beta+\gamma+\delta+\epsilon+\zeta+\eta = 128 \bigr\}.
\label{eq:composition_ext}
\end{align}
The ground truth values for bulk modulus and alloy formation energy were obtained from the converged DFT dataset established by Barry et al.\cite{barry_voxelized_2023, barry_voxelized_2020}. Note that the alloy formation energies were available only for $\mathcal{D}_4$.

\section*{Background}

\subsection*{Pseudo-Density}

In this framework, the material structure is defined by the pseudo-density field, $\rho_{\text{pseudo}}:\Omega\rightarrow\mathbb{R}_{\geq0}$, where $\Omega\subset\mathbb{R}^3$ represents the spatial domain of the atomic system. This framework's key innovation is bypassing the computationally expensive SCF cycle. While converged densities require iterative Hamiltonian diagonalization, pseudo-densities are constructed via a single-shot superposition of isolated atom electron densities \cite{certik_dftatom_2013, bhowmik_spectral_2025}, where each element contributes its valence electron density according to its specific spatial distribution. This effectively decouples feature generation from the computationally expensive electronic relaxation process.

We modeled the random solid solution behavior of refractory HEAs through Special Quasirandom Structures (SQS)\cite{zunger_special_1990} generated via the Alloy Theoretic Automated Toolkit (ATAT)\cite{van_de_walle_alloy_2002} using \textit{sqsgenerator}\cite{gehringer_models_2023}. All SQS structures were generated with the same set of lattice parameters. For each structure, the pseudo-density was computed using valence electron densities in the Optimized Norm-Conserving Vanderbilt (ONCV) \cite{hamann_optimized_2013} pseudopotential files from the SPMS set\cite{shojaei_soft_2023}. We employed the PBE exchange-correlation functional inherent to these pseudopotentials. While higher-level functionals (e.g., SCAN or Hybrids) offer improved energetic accuracy for ground-state calculations, the non-interacting pseudo-density relies primarily on the spatial topology of valence orbitals rather than absolute energy minimization. The Generalized Gradient Approximation (GGA)\cite{perdew_generalized_1996} sufficiently captures the characteristic atomic radii and overlap features required for this topological mapping, while maintaining consistency with standard high-throughput screening libraries. The pseudo-density was mapped onto a 3D real-space grid closely mimicking the implementation in the M-SPARC\cite{zhang_version_2023, xu_m-sparc_2020} and SPARC\cite{zhang_sparc_2024, xu_sparc_2021} electronic structure codes. Critically, the generation of $\rho_{\text{pseudo}}$ bypasses the iterative SCF procedure\cite{ren_impacts_2022}, hence the computational cost is significantly reduced compared to fully converged DFT calculations, thereby facilitating high-throughput screening of complex composition spaces.

\subsection*{Feature Engineering using Spatial Correlations}
To quantify the salient structural features governing the material response, we employ directionally resolved two-point spatial correlations\cite{kalidindi_hierarchical_2015, fullwood_gradient-based_2008}. This measure captures spatial positioning of charge density features, creating a translationally invariant descriptor. For computational efficiency, the discrete two-point autocorrelations, $f_r$, are calculated using Fast Fourier Transforms (FFT)\cite{fast_formulation_2011, niezgoda_delineation_2008, niezgoda_understanding_2011} on a discretized grid:

\begin{equation}
f_r = \frac{1}{|S|}\mathcal{F}^{-1}\left[\mathcal{F}\left(\sqrt{\rho_{\text{pseudo}}}\right)\ast \mathcal{F}\left(\sqrt{\rho_{\text{pseudo}}}\right)\right] \,,
\end{equation}

\noindent where $\mathcal{F}$ denotes the discrete Fourier transform operation, the asterisk ($*$) indicates the autocorrelation operation, and $S$ represents the set of voxel indices. The spatial domain for $\rho_{pseudo}$ was discretized with a fixed voxel size of $\lambda=0.2$ \AA{}, as done in previous work\cite{barry_voxelized_2023} utilizing converged electron charge densitites. This approach enables efficient computation of spatial correlations for high-throughput screening.

The resulting high-dimensional autocorrelation vectors are projected onto a low-dimensional feature space using Principal Component Analysis (PCA)\cite{mackiewicz_principal_1993}. This projection compresses the feature space while preserving the variance that distinguishes different atomic configurations. 

Because $\rho_\text{pseudo}$ requires no iterative Hamiltonian diagonalization (and takes $\sim$30 seconds per sample), its generation scales as $\mathcal{O}(N_\text{grid})$ per structure, a single-pass operation, compared to $\mathcal{O}(N_\text{SCF} \times N_\text{basis}^3)$ for a converged DFT charge density, yielding a reduction in descriptor generation cost of approximately two orders of magnitude for the 128-atom BCC SQS supercells used in this work\cite{barry_voxelized_2023}.

\subsection*{Dimensionality Reduction using PCA}
\begin{figure*}[h]
    \centering
    \includegraphics[width=0.85\textwidth]{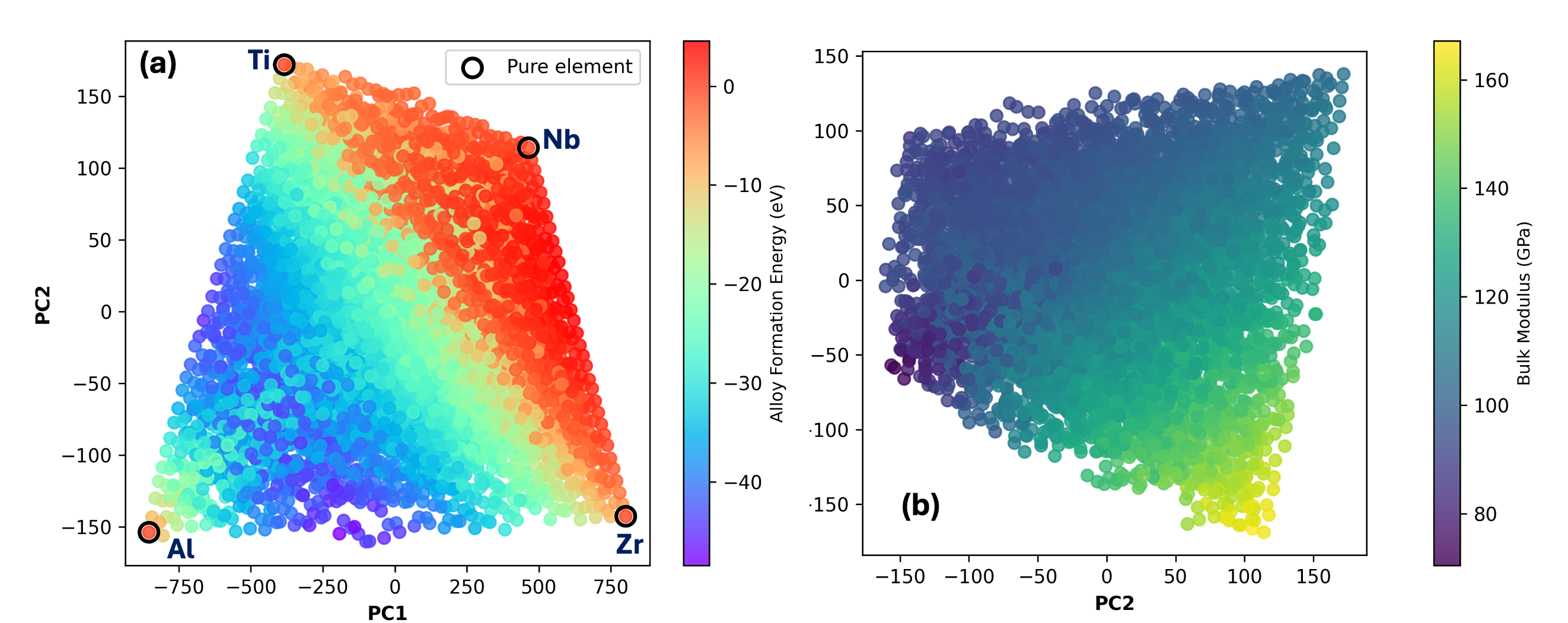}
    \caption{Principal Component analysis of $\mathcal{D}_4$ under a basis estimated from $\mathcal{D}_4$ alone. (a) Alloy Formation Energy and (b) Bulk Modulus, projected onto the first two PCs of a PCA fitted exclusively to the $\mathcal{D}_4$ feature vectors. The trapezoidal simplex topology with pure elements at vertices and alloy compositions filling the interior, is fully preserved under this independent basis, confirming that the compositional manifold structure is an intrinsic property of the pseudo-density representation and not an artifact of the joint $\mathcal{D}_4 + \mathcal{D}_7$ PCA fitting used in the main analysis. Properties vary smoothly and monotonically across the simplex, consistent with composition-dominated variance in this alloy family.}
    \label{fig:PC4only}
\end{figure*}

We apply PCA to the autocorrelation feature vectors to obtain a compact, low-dimensional representation. The PCA transformation matrix is estimated jointly from the combined $\mathcal{D}_4+\mathcal{D}_7$ feature space, ensuring that both datasets are projected onto a common basis. We note that this means the PC-space overlap between $\mathcal{D}_4$ and $\mathcal{D}_7$ (Supplementary Figure \ref{fig:supp1}) reflects partly the shared basis; the physical interpretation that the pseudo-density feature scales of the two alloy families are compatible, is confirmed independently by Supplementary Figure \ref{fig:PC4only}, which shows that the trapezoidal $\mathcal{D}_4$ simplex topology is fully preserved under a PCA estimated from $\mathcal{D}_4$ alone. In the combined feature space, the first three principal components capture approximately 50\% of the total variance, with the first component accounting for approximately 36\%. The PCA basis vectors systematically decompose the structural hierarchy: the first principal component captures the mean charge density, while subsequent components resolve the complex spatial patterns associated with local atomic disorder.

\begin{figure}[!t]
\centering
\includegraphics[width=0.5\textwidth]{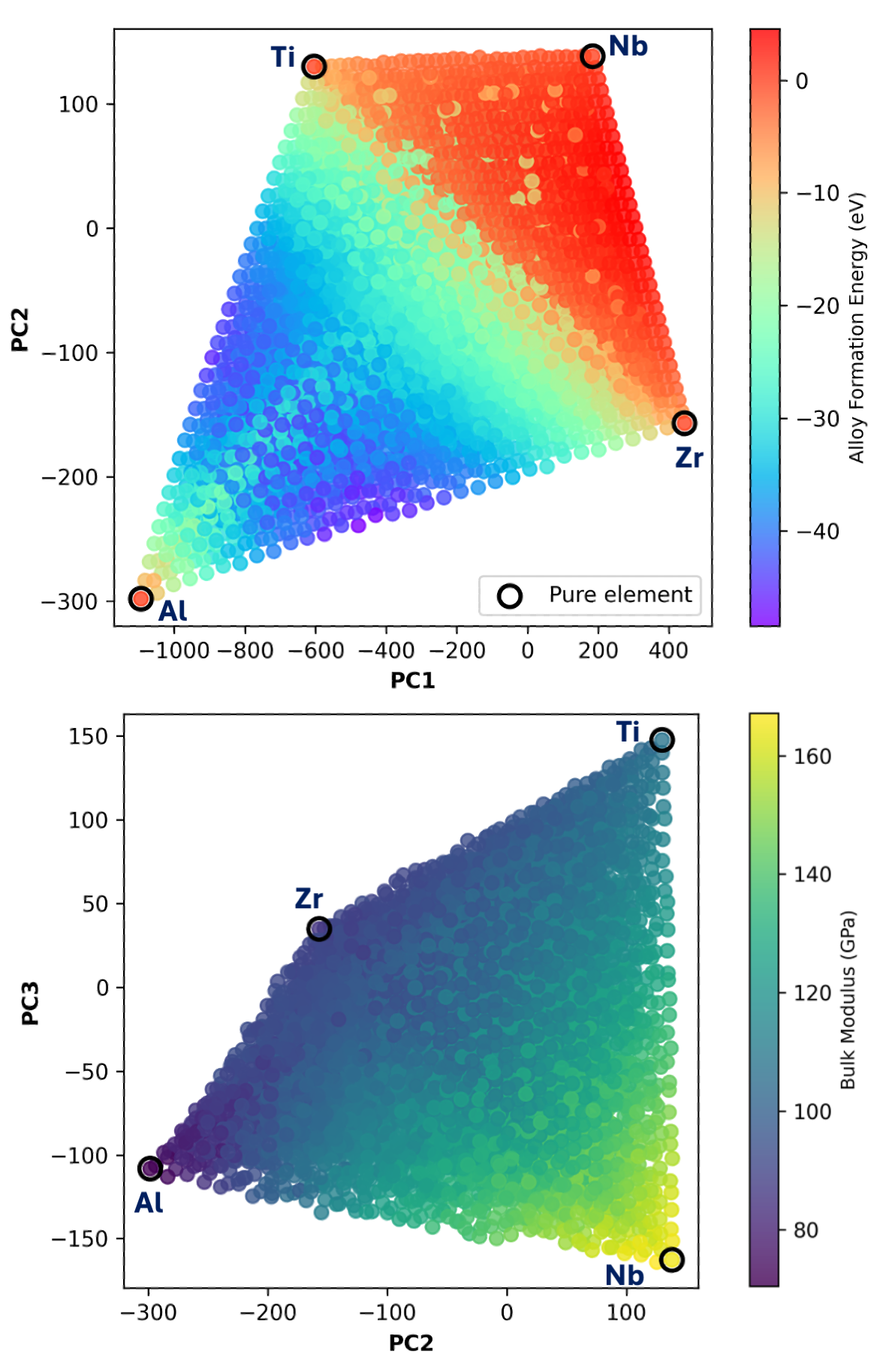}
\caption{\label{fig:PC} Principal component visualization of the $\mathcal{D}_4$ composition space. (top) PC1 vs PC2 colored by alloy formation energy; (bottom) PC2 vs PC3 colored by bulk modulus. Pure elements reside at the vertices of the trapezoidal simplex; alloy compositions fill the interior. Properties vary smoothly and monotonically across the manifold, confirming that the pseudo-density spatial autocorrelations preserve the chemical and structural hierarchy of the composition space. Al occupies a spatially distinct vertex attributable to its s-p valence character, while the d-metal vertices (Ti, Nb, Zr) cluster more closely, consistent with their shared group 4-5 d-electron profiles.}
\end{figure}

\begin{figure*}[h]
    \centering
    \includegraphics[width=0.8\textwidth]{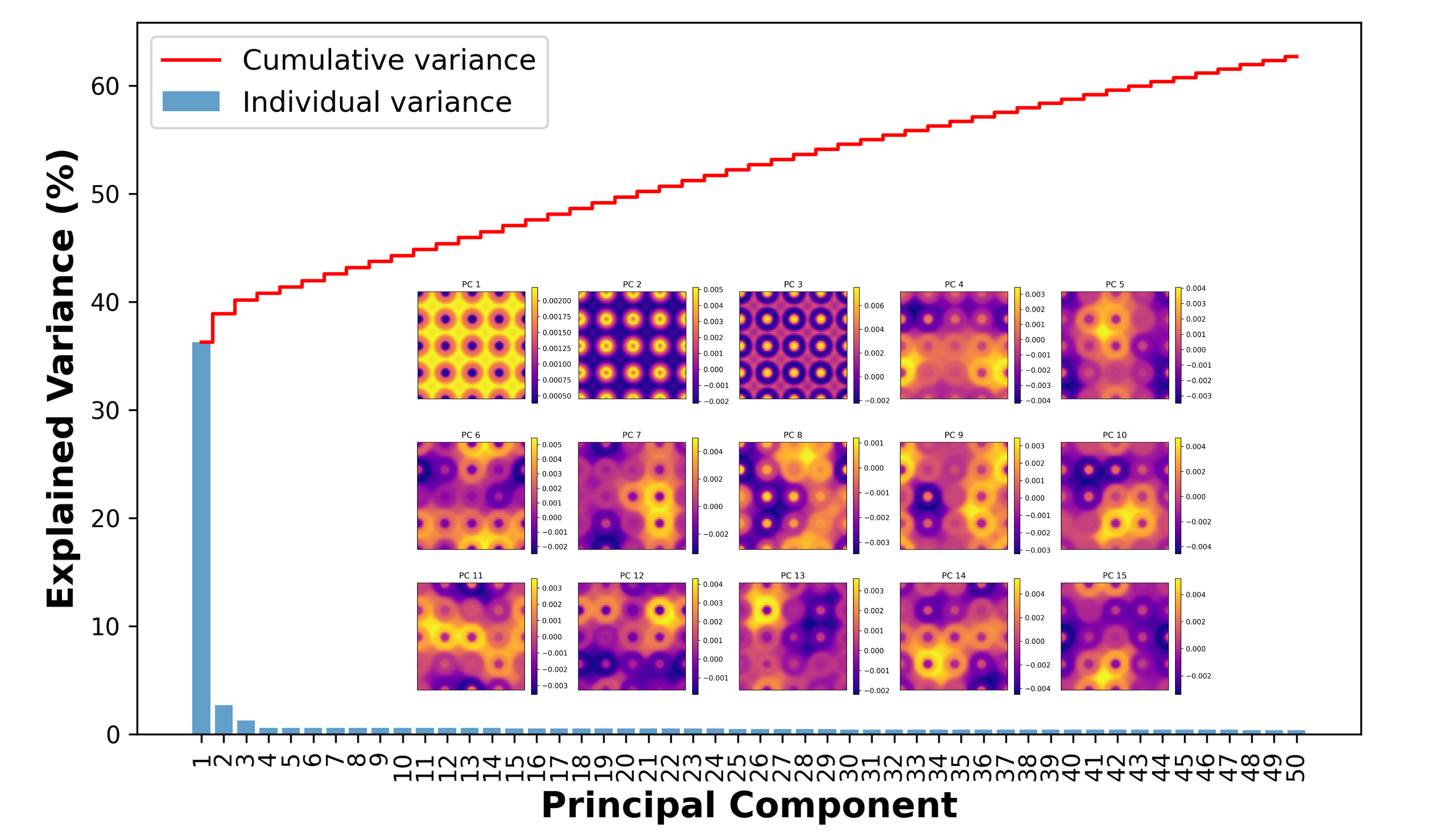}
    \caption{Variance decomposition and loading-vector analysis of the pseudo-density PCA. (top) Scree plot showing the individual explained variance ratio (bars) and cumulative explained variance (red line) for the first 50 principal components of the combined $\mathcal{D}_4 + \mathcal{D}_7$ feature space. PC1 captures $\sim$36\% of total variance; PCs 2 and 3 capture $\sim$2.5\% and $\sim$2\% respectively; PCs 4--50 each contribute $<$0.5\%, producing a near-flat decay with no discernible elbow beyond PC3. (bottom panels) Center slices of the 3D autocorrelation loading vectors for PC1--15. PC1--3 display spatially coherent, periodic patterns consistent with the BCC lattice periodicity, elemental density contrast, and compositional disorder modulations respectively. PC4 and above exhibit progressively higher spatial-frequency content with no periodic lattice structure, indicating that these components encode stochastic variation in the autocorrelation representation rather than chemically interpretable structural features. The combination of the scree-plot elbow and the qualitative transition in loading-vector spatial coherence provides the dual justification for truncating the descriptor space at three principal components.}
    \label{fig:PC7scree}
\end{figure*}

The low-dimensional representation maintains the physical hierarchy of the alloy system. As illustrated in Figure \ref{fig:PC} for $\mathcal{D}_4$, the data points naturally arrange into a trapezoidal geometry where pure elements occupy the vertices and all alloy compositions fill the interior volume. This smooth variation of principal component scores across the composition space facilitates robust interpolation and accurate property prediction for unexplored alloy stoichiometries. Because this dimensionality reduction is unsupervised, the resulting descriptors are property-agnostic and can be reused to train independent models for different target properties, e.g., mechanical or thermodynamic, without re-computation. Notably, this unsupervised PC projection remains fixed regardless of the target property, allowing the same structural map to be colored by distinct physical responses, e.g., bulk modulus or alloy formation energy: see Figure \ref{fig:PC} to reveal property-specific manifolds. Based on the variance and loading-vector analysis, we truncate the descriptor space to the first three principal components \footnote{see Supplementary Figure \ref{fig:manifold_topology} (a) and (b) and \ref{fig:PC7scree} (scree plot and PC basis vectors)} for all subsequent regression modeling. The scree plot shows a pronounced elbow after PC3, with PC1 capturing $\sim$36\% of variance and PCs 2-3 capturing $\sim$2.5\% and $\sim$2\% respectively, followed by a near-flat decay through PC4--50 (each $<$0.5\%). Inspection of the PC loading vectors (center slices of the 3D autocorrelation basis) confirms that PC1--3 display spatially coherent, periodic patterns reflecting the BCC lattice structure, elemental density contrast, and compositional disorder respectively. PC4 and above exhibit progressively higher spatial-frequency content with no discernible periodic structure, indicating that they encode stochastic noise in the autocorrelation representation rather than chemically interpretable structural features. These three components therefore capture the full chemically relevant structural hierarchy, and their empirical sufficiency is further evidenced by $R^2 \geq 0.98$ achieved in-domain.

\subsection*{Impact of Structural Relaxation on Manifold Topology}

A critical and counter-intuitive finding of this work is that structural descriptors derived from initial SQS configurations with a uniform, common lattice constant yield a more cohesive feature space than those derived from fully relaxed geometries. Figure \ref{fig:PC} visualizes the Principal Component space for the initial uniform SQS structures compared to those relaxed via VASP\cite{kresse_ab_1993, kresse_efficiency_1996, kresse_efficient_1996, kresse_ultrasoft_1999} and MACE\cite{maes_mace_2025} (see plots c, d, e and f in Supplementary Figure \ref{fig:manifold_topology}). Here, relaxation refers to the full optimization of both internal ionic coordinates and cell volume. The uniform SQS data forms a continuous, cohesive simplex, where the variance is driven purely by the combinatorial arrangement of chemical species. In contrast, the relaxed structures fracture into a distinct, disjoint, hyper-branched topology. 

This fracturing is fundamentally driven by the loss of volumetric uniformity during relaxation. In the uniform initial state, a globally constant spatial metric is enforced for the two-point spatial statistics across all samples. During relaxation, individual lattice parameters deviate significantly from this uniform baseline, rendering the spatial metric sample-dependent. As the concentration of larger atoms like Zr increases (Zr's ionic radius is approximately 10\% larger than that of Al, Nb, and Ti), the lattice undergoes significant global cell expansion to minimize steric hindrance.

\begin{figure*}[h]
    \centering
    \includegraphics[width=0.7\textwidth]{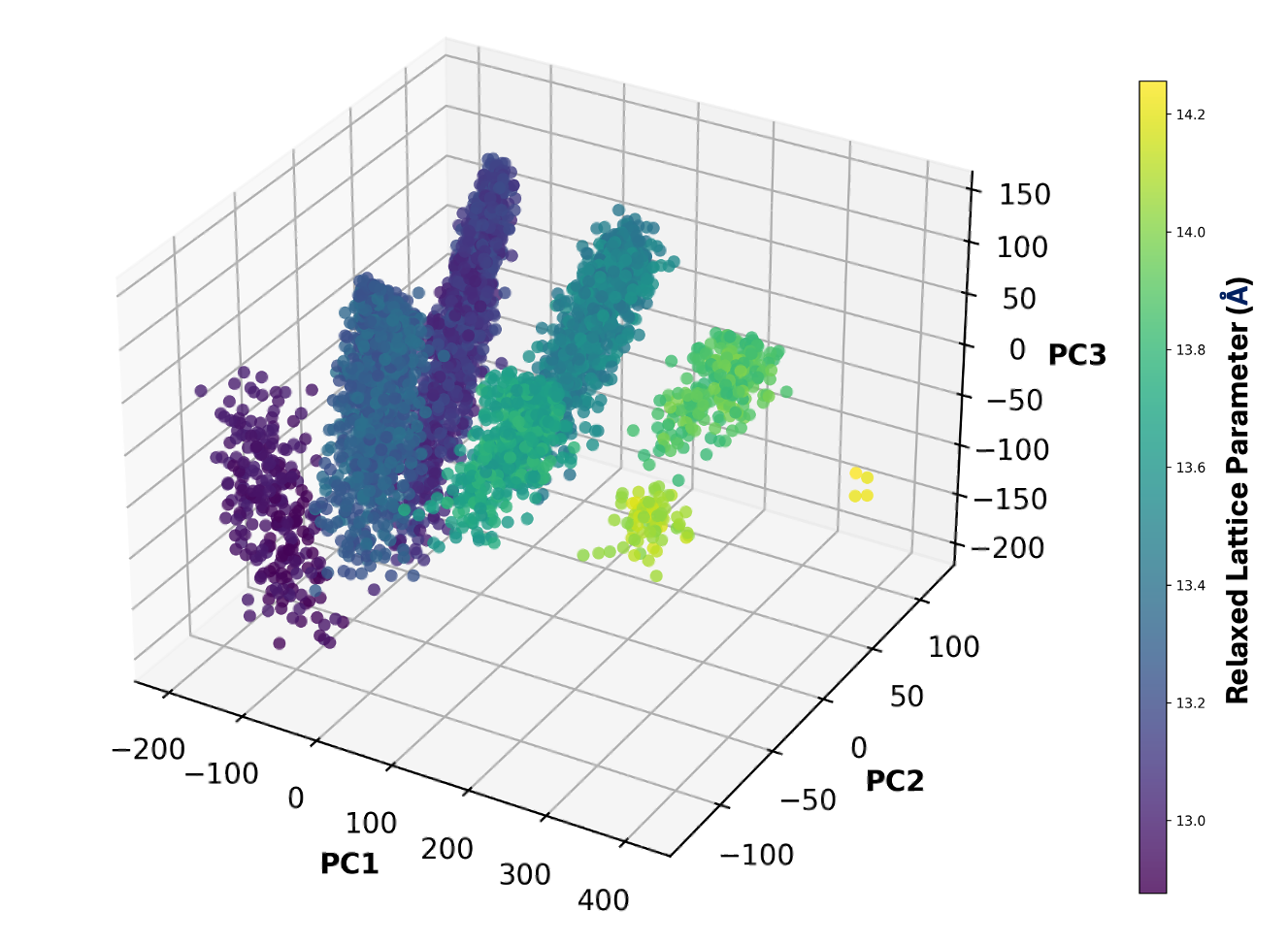}
    \caption{\textbf{Stratification of the relaxed manifold by cell volume. (shown $\mathcal{D}_4$ quaternary samples)} The Principal Component space of the fully relaxed dataset colored by the final relaxed lattice parameter. The aggressive variance maximization of PCA captures the spatial dilation of the absolute-grid $f_r$, perfectly stratifying the disjoint branches by discrete lattice parameter bands ranging from 12.8 \AA\ to 14.2 \AA.}
    \label{fig:relaxed_colored_lp}
\end{figure*}

This phenomenon can be rigorously understood by considering the variance maximization objective of PCA acting on absolute spatial coordinates. Because the $f_r$ are computed on an absolute real-space grid, global cell expansion physically dilates the interatomic distances between charge density peaks. Consequently, the variance is dominated by different lattice constants scaling the spatial metric, rather than strictly chemical variations. PCA aggressively captures this spatial dilation, fracturing the manifold into discrete arms. When the relaxed PC space is colored by the final relaxed lattice parameter, the distinct branches stratify perfectly by cell volume, with the arms corresponding to discrete lattice parameter bands (see Figure \ref{fig:relaxed_colored_lp}) scaling from approximately 12.8 \AA\ to 14.2 \AA. The terminal tips of these disjoint manifolds correspond exactly to the absolute structural limits of the dataset, specifically the highly expanded Zr-rich compositions. 

\begin{figure*}[h]
    \centering
    \includegraphics[width=0.8\textwidth]{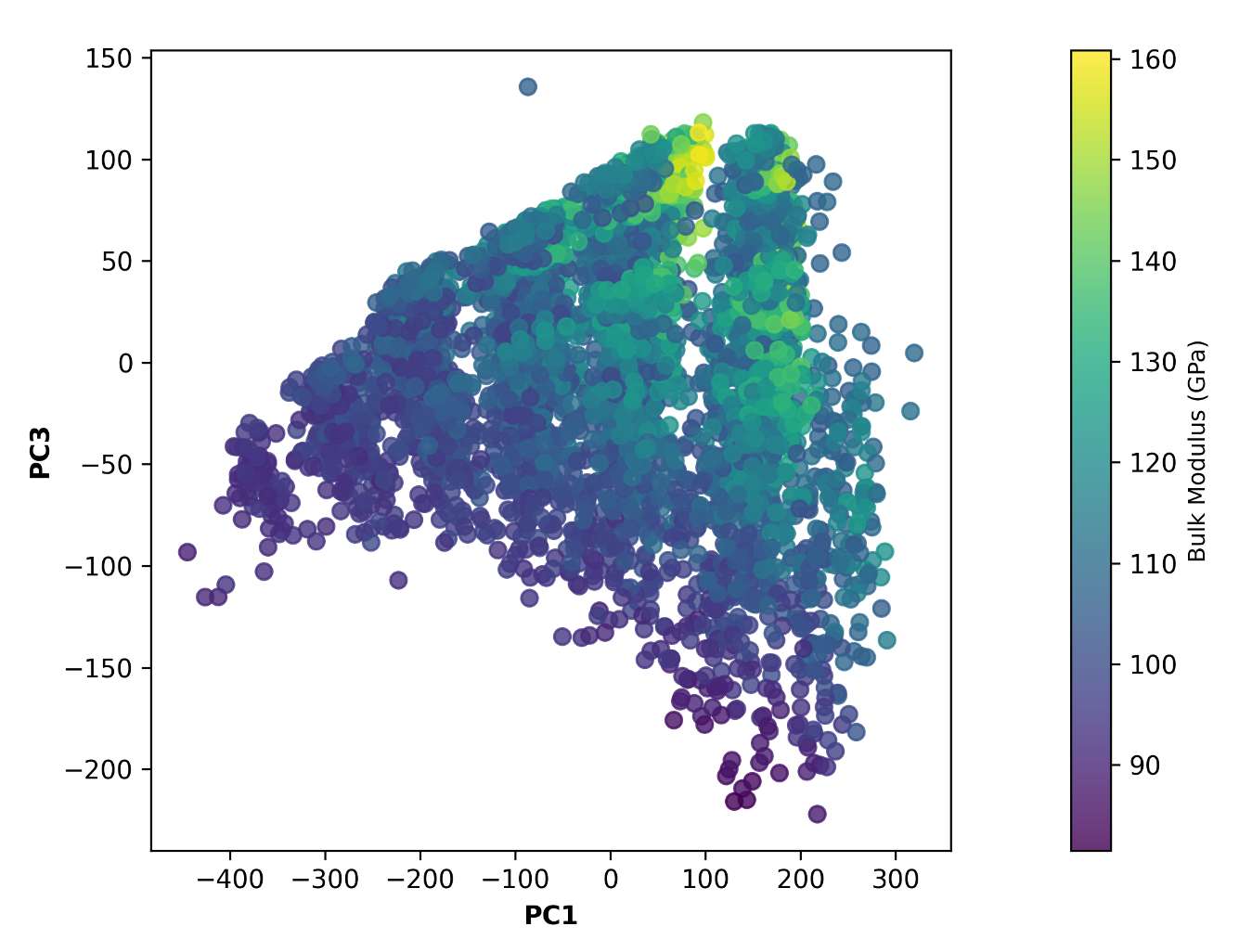} 
    \caption{\textbf{Validation of volumetric scaling effects (shown $\mathcal{D}_4$ quaternary samples).} When the absolute-grid 2-point spatial statistics ($f_r$) of the relaxed structures are divided by the cell volume ($a^3$) prior to PCA, the magnitude-dependent variance driven by global cell expansion is normalized. This suppression of volume-dependent information causes the previously disjoint, hyper-branched manifold to coalesce into a single, continuous cluster, confirming that non-uniform final lattice constants are the primary driver of the topological fracturing.}
    \label{fig:coalesced_manifold}
\end{figure*}

To rigorously validate that this non-uniform volumetric scaling is the primary driver of the fracturing, we performed an ablation test where the absolute-grid $f_r$ were divided by the cell volume ($a^3$) before performing PCA (see Figure \ref{fig:coalesced_manifold}). This operation effectively normalizes the magnitude-dependent variance driven by the cell expansion. Suppressing this volume-dependent information caused the previously disjoint branches to coalesce back into a single, continuous cluster. This mathematically confirms that the geometric relaxation, specifically the variance in the final relaxed lattice constants, is responsible for the manifold fracturing.

It is worth noting that if the $f_r$ were computed on a normalized grid using relative fractional coordinates ($x/a$), the affine volume expansion would be factored out, isolating the variance purely to local ionic displacements. However, by using the initial system's pseudo-density with a uniform lattice constant on an absolute grid, we inherently bypass both the volumetric and ionic strain artifacts. By enforcing this globally constant spatial metric, we effectively treat the ideal lattice as a canonical reference state. It is important to acknowledge the trade-off explicitly: the smoother manifold topology observed for unrelaxed structures arises in part because physically meaningful variability-lattice-mismatch strain, local ionic distortion, and site-specific bonding has been suppressed. For macroscopic, composition-averaged properties of ideal solid solutions, where bulk modulus and formation enthalpy are primarily determined by elemental identity and proportion rather than by individual SQS configurations, this suppression is a defensible approximation. The framework is not designed to capture configuration-sensitive properties, and performance on such quantities has not been tested. This enables the Gaussian Process\cite{rasmussen_gaussian_2005} to interpolate across a smooth manifold ($f(\text{chemistry}) \rightarrow P_{\text{property}}$) without the interference of high-variance geometric noise introduced by relaxation.

\subsubsection*{Physical basis for descriptor efficacy and transferability}
Several physically grounded arguments underpin the efficacy of the pseudo-density for the properties and alloy class studied here. First, for macroscopic, composition-averaged properties of ideal disordered BCC solid solutions, such as bulk modulus and formation enthalpy, the dominant source of inter-composition variance is elemental identity and proportion rather than individual atomic configurations. In this regime, a descriptor that faithfully encodes the spatial envelope of valence electron overlap captures the primary variance driver without requiring full self-consistency. Second, the omission of electron-electron interactions and SCF relaxation is acceptable for these specific properties because electronic screening in metallic solid solutions constitutes a subordinate perturbation on the zeroth-order charge topology set by the pseudo-density; the self-consistent correction is a composition-invariant shift that does not alter the relative ordering of compositions across the manifold. This is consistent with the empirical observation that the unrelaxed pseudo-density manifold is smoother and more predictive than that of relaxed structures. Third, intra-family transferability from $\mathcal{D}_4$ to $\mathcal{D}_7$ is physically enabled by the similarity of d-electron radial density profiles across the group 4--6 refractory metals (Ti, Zr, Nb, Mo, Ta, V, W): these elements share the same angular momentum quantum number and comparable effective nuclear charges, so their pseudopotential-derived valence densities are similar in spatial extent and nodal structure. Consequently, the autocorrelation features of Mo, Ta, V, and W fall in a region of feature space continuously connected to those of Nb, Ti, and Zr, enabling regression rather than requiring interpolation between disjoint clusters. Fourth, Al is an s-p metal whose diffuse, nearly-spherical valence density produces a qualitatively distinct autocorrelation signature, placing it at a spatially separated vertex of the $\mathcal{D}_4$ simplex (visible in Figure~\ref{fig:PC}). Because Al is absent from $\mathcal{D}_7$, all $\mathcal{D}_7$ compositions lie in the d-metal-dominated region of the joint manifold, away from the Al vertex. The zero-shot and few-shot predictions for $\mathcal{D}_7$ therefore extrapolate within a chemically coherent subspace and do not require the model to generalize through the Al-anchored region of the descriptor space. It must be emphasized that this framework is explicitly designed for composition-averaged properties of single-phase disordered BCC solid solutions and is not expected to capture properties governed by local chemical order, short-range order, segregation, magnetic ordering, defects, or finite-temperature configurational sampling.

\subsection*{Model Building}

\subsubsection*{Gaussian Process Regression}
We employ Gaussian Process Regression (GPR)\cite{rasmussen_gaussian_2010} to model the mapping between the low-dimensional structural features (PC scores) and the material properties. GPR provides a non-parametric, probabilistic framework that yields both a predictive mean and a variance, which is essential for uncertainty quantification. The covariance between inputs is computed using an Automatic Relevance Determination Squared Exponential (ARDSE)\cite{rasmussen_gaussian_2006} kernel:

\begin{equation}
k(\mathbf{x},\mathbf{x}') = \sigma_s^2\exp\left[-\frac{1}{2}\sum_{d=1}^{D}\frac{(x_d-x'_d)^2}{l_d^2}\right] + \sigma_n^2\delta_{\mathbf{x}\mathbf{x}'}
\end{equation}

\noindent where $\sigma_s^2$ scales the output variance, $l_d$ represents the characteristic length-scale for feature dimension $d$, and $\sigma_n^2$ accounts for observation noise. The ARDSE kernel allows the model to inherently determine the relevance of each principal component, weighting them according to their influence on the target property. This choice reflects the expectation that different principal components encode distinct physical scales, e.g., mean density versus local disorder, and allows the GP to adapt its sensitivity accordingly. All hyperparameters are optimised by maximising the log marginal likelihood using the Adam\cite{kingma_adam_2017} optimiser for 200 epochs at a learning rate of 0.1.

\begin{figure}[h]
\centering
\includegraphics[width=0.5\textwidth]{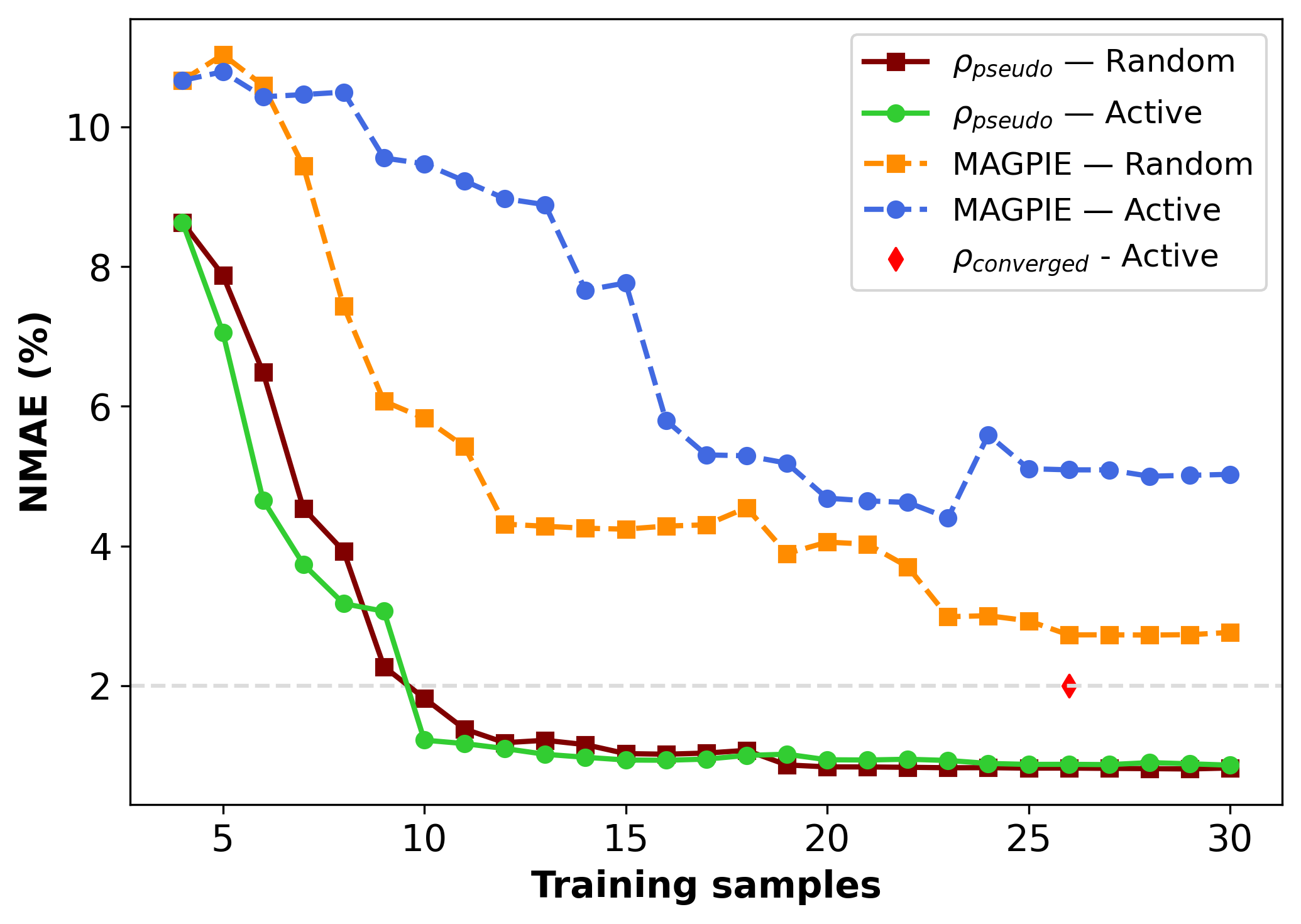}
\caption{\label{fig:bulk_conv}  Controlled comparison of bulk modulus prediction error (NMAE) as a function of training set size on $\mathcal{D}_4$, for three descriptor classes under the identical GPR and active-learning pipeline: pseudo-density spatial autocorrelations ($\rho_\text{pseudo}$, active and random), composition-based features (active and random), and a single reference datum for the converged charge-density descriptor ($\rho_\text{converged}$, active, at $\sim$26 samples\cite{barry_voxelized_2023}). The pseudo-density active strategy reaches NMAE $<$2\% at 10 samples. The composition-based descriptor does not reach the 2\% threshold within the 30-sample window evaluated here, asymptoting above 3.5\%, confirming that the spatial autocorrelation encoding of the valence electron density provides structural information beyond elemental composition statistics alone.}
\end{figure}

\begin{figure*}[ht!]
\centering
\includegraphics[width=\textwidth]{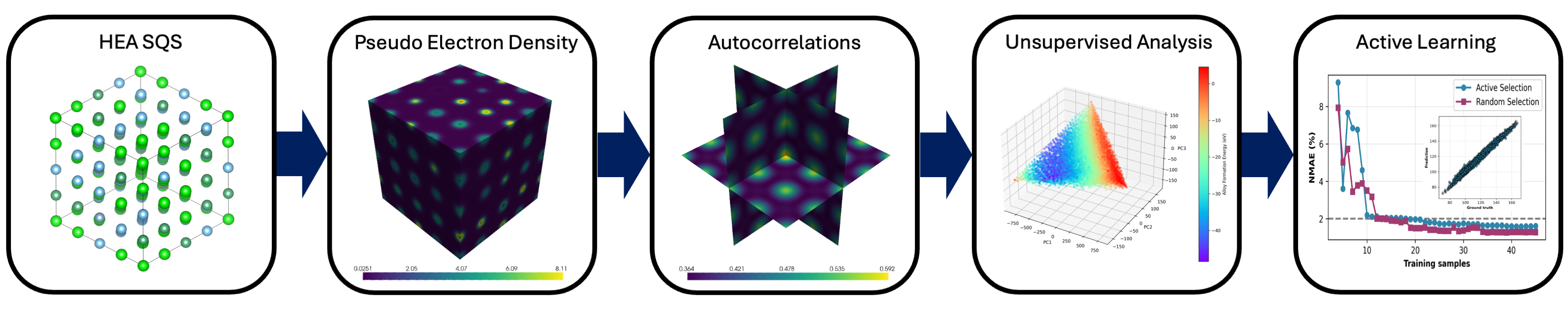}
\caption{\label{fig:workflow} End-to-end Bayesian active learning workflow for $\mathcal{D}_4$. Starting from HEA Special Quasirandom Structures (SQS), the pseudo-density $\rho_\text{pseudo}$ is constructed via single-pass superposition of isolated valence densities, bypassing the SCF cycle. Two-point spatial autocorrelations $f_r$ are computed via FFT and compressed to three principal components. The GPR model is iteratively updated by querying ground-truth DFT labels only at the highest-uncertainty candidates identified by the acquisition function $I(\mathbf{x}) = |\sigma(\mathbf{x})/\mu(\mathbf{x})|$, minimizing the number of expensive DFT calculations required.}
\end{figure*}

\subsubsection*{Bayesian Experiment Design}
To minimize the computational expense of data generation, we utilize a Bayesian active learning strategy driven by a relative-uncertainty acquisition function inspired by information-based design\cite{lindley_measure_1956, huan_simulation-based_2013}. Specifically, the acquisition selects candidate structures that maximize the ratio of predictive uncertainty to the predicted magnitude:
\begin{equation}
I(\mathbf{x}) = \left|\frac{\sigma(\mathbf{x})}{\mu(\mathbf{x})}\right|
\end{equation}
where $\mu(\mathbf{x})$ and $\sigma(\mathbf{x})$ are the predictive mean and standard deviation provided by the GPR model, respectively. In each iteration, the algorithm identifies the $k$ structure with the highest $I(\mathbf{x})$ for ground-truth evaluation. In a prospective discovery campaign, this step would selectively trigger DFT calculations for these specific unlabelled candidates, thereby augmenting the training set with high-fidelity data only where strictly necessary. While active learning strategies have been successfully applied to descriptors derived from converged densities\cite{barry_voxelized_2023}, such workflows inherently face a "pre-computation" bottleneck: the computationally expensive SCF cycle must be completed for every candidate structure merely to generate the input features for the surrogate model. In contrast, our pseudo-density framework eliminates this redundancy, enabling the rapid, low-cost scanning of the entire candidate pool prior to triggering any expensive DFT calculations. This iterative process (see Figure \ref{fig:workflow}) empirically leads to rapid convergence by prioritizing sampling in regions in the design space where the predictive confidence is lowest.

\subsubsection*{Extrapolative Validation Protocol}
To rigorously assess the transferability of the pseudo-density descriptors, we designed a disjoint training-testing protocol. The GPR model was actively trained exclusively on samples from the lower-order domain $\mathcal{D}_4$ (Al-Nb-Ti-Zr). This model was then frozen and tasked with predicting the properties of the full extrapolation domain $\mathcal{D}_7$ (Mo-Nb-Ta-Ti-V-W-Zr) without any re-training or exposure to the new chemical elements. This zero-shot transfer protocol evaluates whether the learned regression mapping $f: \mathbf{PC} \rightarrow K$ generalizes across chemically distinct refractory systems without re-training, fine-tuning, or element-specific feature engineering. Both $\mathcal{D}_4$ and $\mathcal{D}_7$ share the BCC crystal structure and consist exclusively of group 4 to 6 transition metals; the transfer demonstrated here is therefore intra-family transfer within the refractory BCC alloy class, not unconditional extrapolation to arbitrary chemistries. Successful transfer implies that the pseudo-density encodes valence-overlap features that are transferable within this structurally and chemically constrained family.

\subsection*{Error metrics}

Across all experiments, model performance is quantified using the mean absolute error (MAE), normalized MAE (NMAE), mean absolute percentage error (MAPE), and coefficient of determination ($R^2$). For a set of $N$ predictions $\{\hat{y}_i\}$ and corresponding ground-truth values $\{y_i\}$, these are defined as
\begin{align}
\mathrm{MAE} &= \frac{1}{N} \sum_{i=1}^N \left| \hat{y}_i - y_i \right|, \\
\mathrm{NMAE} &= \frac{\mathrm{MAE}}{\bar{y}} \times 100\%,
\end{align}
where $\bar{y}$ is the mean of the ground-truth values $\{y_i\}$.

\section*{Results and Discussion}

\begin{figure*}[t!]
\centering
\includegraphics[width=0.8\textwidth]{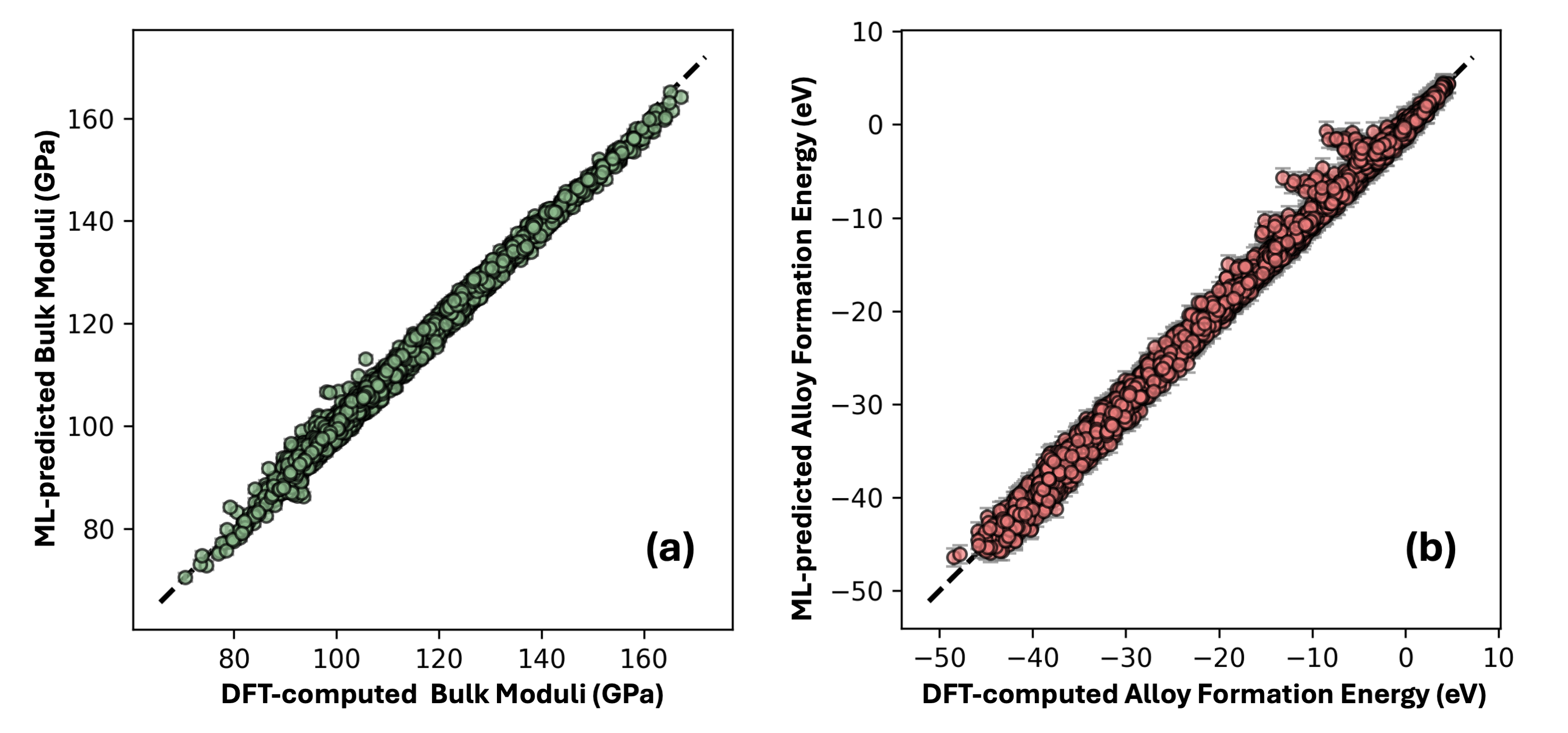}
\caption{\label{fig:bulk_parity} Parity plots of predicted versus DFT-computed properties for all held-out samples in $\mathcal{D}_4$, with GPR predictive uncertainty shown as error bars ($\pm 1\sigma$). (a) Bulk modulus predictions using a model trained on 10 actively selected samples ($R^2 = 0.98$, NMAE = 1.55\%). (b) Alloy formation energy predictions using a distinct GPR model trained on 18 actively selected samples ($R^2 = 0.99$, NMAE = 2.20\%), with uncertainty coverage of 88.8\% within $\pm 1\sigma$ and 99.8\% within $\pm 2\sigma$.}
\end{figure*}

\begin{figure*}[ht!]
\centering
\includegraphics[width=0.8\textwidth]{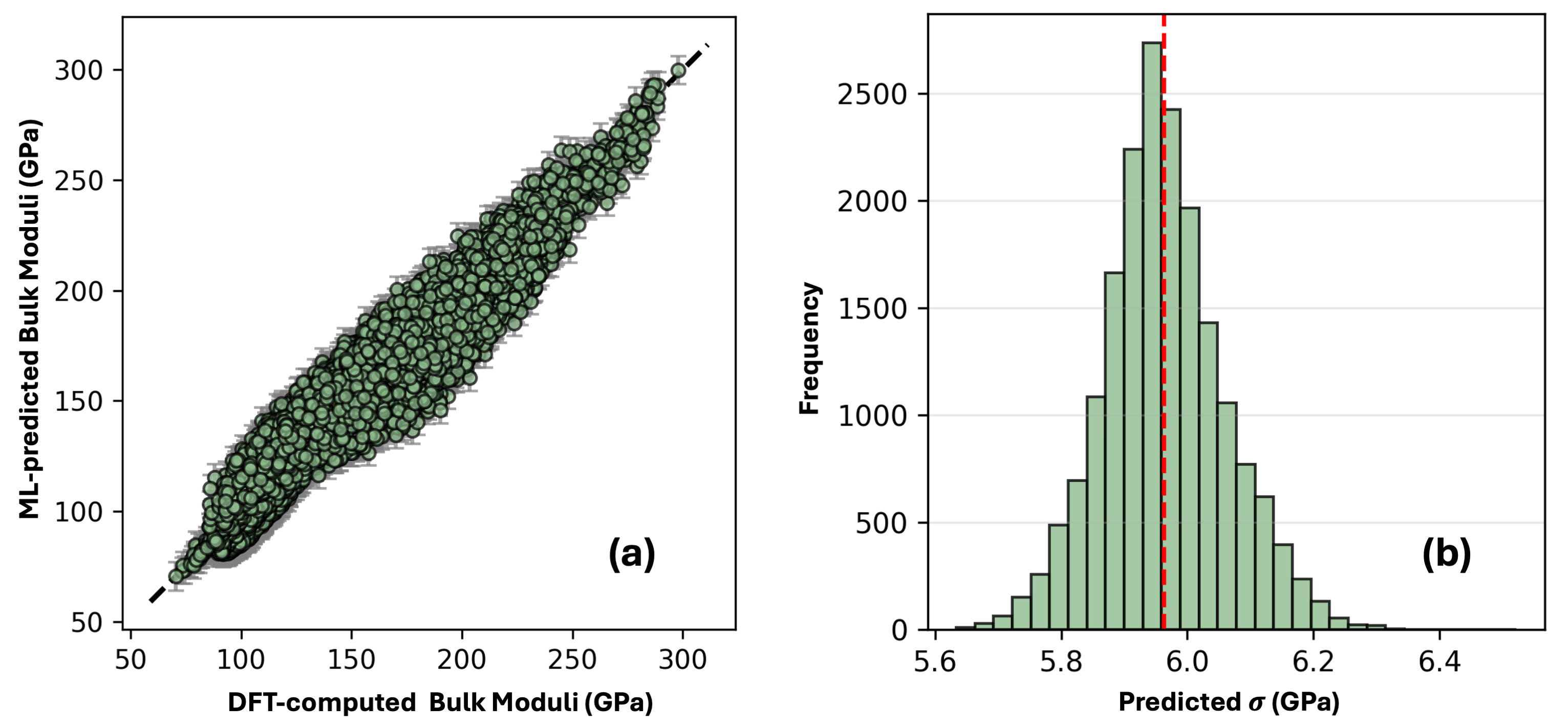}
\caption{\label{fig:extrapolation_parity} Extrapolative performance of the $\mathcal{D}_4$-trained model on the full $\mathcal{D}_7$ bulk modulus dataset (12,012 compositions), following few-shot augmentation with 20 actively selected $\mathcal{D}_7$ samples. (a) Parity plot with GPR predictive uncertainty shown as error bars ($\pm 1\sigma$). (b) Distribution of predicted standard deviations $\sigma$ across all $\mathcal{D}_7$ test compositions. Coverage of 43.2\% within $\pm 1\sigma$ and 78.8\% within $\pm 2\sigma$ indicates moderate overconfidence in the extrapolative regime, consistent with GP posterior uncertainty being prior-dominated outside the training chemical space.}
\end{figure*}

\subsection*{Performance on Bulk Modulus predictions}
The GPR model demonstrates high predictive accuracy for the bulk modulus across the complete compositional dataset ($\mathcal{D}_4$). The active learning campaign selects labels from a pre-existing DFT candidate pool over a fixed compositional grid\cite{barry_voxelized_2023}, initialized with a minimal seed of 4 randomly selected samples, followed by iterative acquisition. As shown in Figure \ref{fig:bulk_conv}, the model stabilizes rapidly, achieving an NMAE of \textless2\% with only 10 total training samples (4 seed + 6 active). This efficiency highlights a significant advantage over competing descriptor frameworks. While state-of-the-art methods typically require significantly larger datasets ($10^2$–$10^3$ samples) or approximately 26 samples for converged charge density descriptors\cite{barry_voxelized_2023} to reach convergence, our approach achieves this fidelity with fewer samples and orders of magnitude lower computational cost for feature generation.

The robustness of the pseudo-density descriptor is further evidenced by the performance of random sampling. As summarized in Table \ref{tab:all}, even the random selection strategy yields high accuracy ($R^2 = 0.98$, MAE = 1.41 GPa) on the full dataset ($\mathcal{D}_4$). This indicates that the unrelaxed pseudo-density manifold is naturally well-correlated with the mechanical response, such that complex active learning acquisition functions are not strictly required to achieve good global accuracy; though they still offer superior efficiency in the low-data limit (Figure \ref{fig:bulk_conv}). The parity plot in Figure \ref{fig:bulk_parity}(a) confirms this strong linear correlation ($R^2 = 0.98$) across the full range of 80 GPa to 160 GPa, confirming that the superposition of non-interacting electron densities contains sufficient physical information to resolve variations in mechanical stiffness without systematic bias. Uncertainty calibration on the held-out $\mathcal{D}_4$ bulk modulus predictions yields coverage of 65.1\% within $\pm 1\sigma$ and 92.3\% within $\pm 2\sigma$, consistent with a well-calibrated probabilistic model near theoretical Gaussian expectations (68\%, 95\%).

\begin{table}[h]
\small
\centering
\caption{\label{tab:all} Compiled error metrics for bulk modulus and alloy formation energy predictions for all $\mathcal{D}_4$ compositions (6,545 structures).}
\begin{tabular*}{\textwidth}{lcccc}
\cmidrule(l{0em}r{-13em}){1-2}
Material Property & Strategy & MAE & NMAE (\%) & $R^2$ \\
\cmidrule(l{0em}r{-13em}){1-2}
\multirow{2}{*}{Bulk Modulus (GPa)} & Active & 1.50 & 1.55 & 0.98 \\
 & Random & 1.41 & 1.46 & 0.98 \\
\multirow{2}{*}{Alloy Formation Energy (eV)} & Active & 1.19 & 2.20 & 0.99 \\
 & Random & 1.23 & 2.33 & 0.97 \\
\cmidrule(l{0em}r{-13em}){1-2}
\end{tabular*}
\end{table}

\subsection*{Performance on Alloy Formation Energy predictions}
Crucially, this physics-based feature set exhibits high transferability across properties. The same 3-PC pseudo-density features accurately predict both mechanical stiffness (bulk modulus) and thermodynamic stability (alloy formation energy) without modification. Starting with the same initial seed size of 4 samples, the active selection strategy applied to the full compositional space of $\mathcal{D}_4$ (Table \ref{tab:all}), achieves an MAE of $1.19$ eV (2.20\% NMAE) with an $R^2$ of 0.99.

The parity plot in Figure \ref{fig:bulk_parity}(b) confirms this strong correlation using only 18 actively selected training samples. This result indicates that the pseudo electron density spatial correlations effectively capture the local chemical environment variations that dictate the energetic stability of the solid solution. The uncertainty coverage for $\Delta E_{\text{form}}$ is similarly robust, achieving $88.8\%$ within $\pm 1\sigma$ and $99.8\%$ within $\pm 2\sigma$. 

\subsection*{Comparative Performance and Scientific Implications}

Table \ref{tab:all} summarizes the compiled error metrics, confirming that the pseudo-density descriptor yields high fidelity ($R^2 \geq 0.98$) with both active and random sampling. While both strategies converge to similar accuracy in the limit, Figure \ref{fig:bulk_conv} demonstrates that the active learning strategy provides a superior sample efficiency advantage in the low-data regime. The central finding of this study is the demonstration that pseudo electron densities are a sufficient structural descriptor for high-fidelity property prediction in complex alloy systems. PCA projections of the pseudo electron densities naturally recover the trapezoidal geometry of the 4-element structure space (Figure \ref{fig:PC}), placing elements at the vertices and alloys in the interior. This confirms that atomic packing and chemical identity; both captured by the pseudo-density, dominate the structural hierarchy of HEAs. Electronic relaxation acts predominantly as a local perturbation relative to these primary chemical variations, though the geometric variance introduced by ionic relaxation is sufficient to fracture the low-dimensional manifold if not accounted for (as confirmed by both PCA and Partial Least Squares analyses). This indicates that for these HEA solid solutions, the electronic environment is dominated by the initial, superposition of atomic charges, and the subsequent self-consistent electronic relaxation introduces differences that are subordinate to the chemical variations across the design space. By replacing the converged density with the pseudo-density, the framework effectively decouples the feature generation step from the most computationally demanding part of the DFT calculation. This combination of eliminating the SCF bottleneck for feature generation and reducing the required number of training samples substantially enhances the feasibility of high-throughput computational materials discovery for HEAs. This efficiency gain does not merely accelerate existing workflows; it unlocks regimes previously inaccessible to DFT-based screening. As demonstrated by the extrapolation results, the pseudo-density descriptor creates a transferable feature space that bridges distinct chemical systems. By projecting the 4-component and 7-component datasets into the same Principal Component space (as discussed in the Methods), we observe that they occupy overlapping manifolds. This indicates that the model predominantly learns electronic packing rules rather than overfitting to element-specific labels, enabling more rapid exploration of combinatorial spaces (including 5+ component systems) without the need to retrain on every new element. The pseudo-density's advantage over composition-based representations stems from three physically distinct information channels that elemental fractions alone cannot encode: (i) the \textit{spatial} distribution of valence charge, captured via the two-point autocorrelation $f_r$, which distinguishes arrangements that are compositionally identical but structurally distinct; (ii) the \textit{configurational disorder} of the SQS lattice, whose specific atomic arrangement modulates the local overlap topology of the superimposed densities beyond what is recoverable from mean elemental proportions; and (iii) \textit{electronic character} encoded in the pseudopotential-derived valence densities, which carry element-specific radial and angular orbital information beyond atomic number or mass. To isolate the contribution of the descriptor itself, Figure~\ref{fig:bulk_conv} presents a controlled substitution experiment in which a standard composition-based featurizer is replaced as the descriptor into the \textit{identical} GPR framework and active-learning pipeline. Under this substitution, the composition-based descriptor does not reach the 2\% NMAE threshold within 30 training samples, whereas the pseudo-density active strategy crosses this threshold at 10 samples (converged charge-density descriptors\cite{barry_voxelized_2023} reach comparable accuracy at $\sim$26 samples but at orders-of-magnitude greater computational cost per feature evaluation, as discussed in the Pseudo-Density subsection). This controlled comparison, with all modeling choices held fixed, confirms that the spatial autocorrelation encoding of the pseudo-density provides information beyond elemental composition statistics. Furthermore, composition-based representations are defined over a fixed elemental vocabulary and have no principled mechanism for generalizing to elements absent from training, which is the defining limitation for $\mathcal{D}_4 \rightarrow \mathcal{D}_7$ transfer. The capability to achieve high accuracy for both bulk modulus and formation energy with minimal training data using a low-cost descriptor represents a critical step towards realizing true high-throughput screening of the vast compositional space. 
\begin{figure*}[ht!]
\centering
\includegraphics[width=\textwidth]{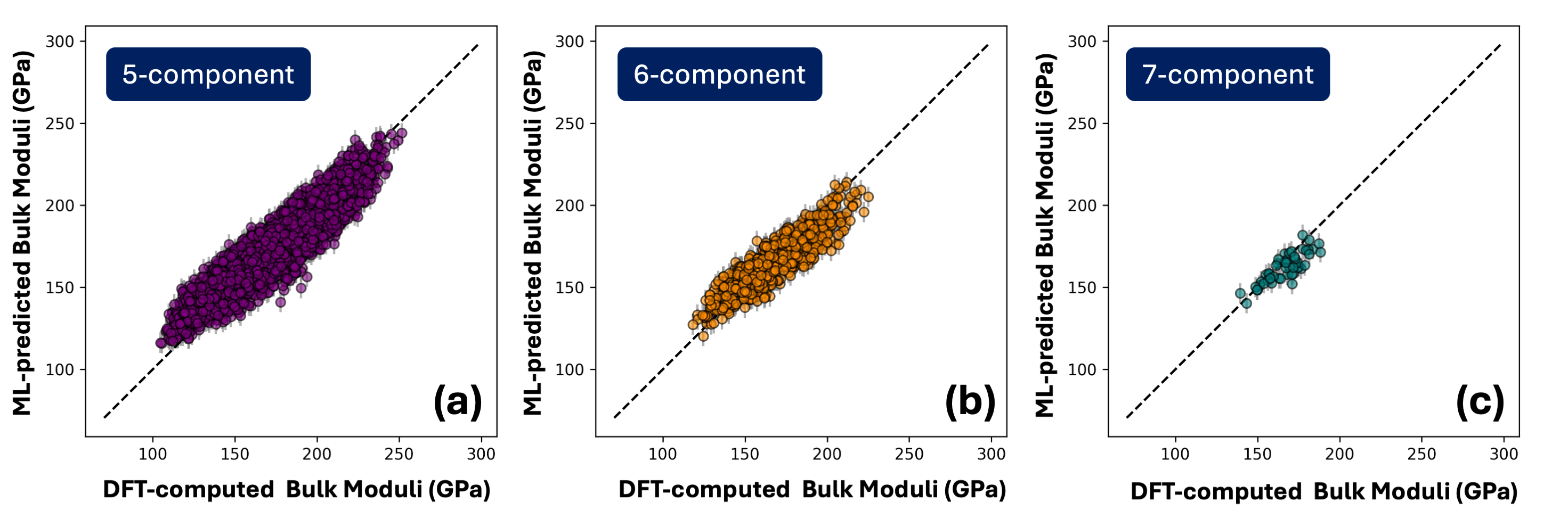}
\caption{\label{fig:parity567} Few-shot domain adaptation performance by chemical complexity within $\mathcal{D}_7$. Parity plots for (a) 5-component ($R^2 = 0.85$, NMAE = 5.54\%), (b) 6-component ($R^2 = 0.82$, NMAE = 4.16\%), and (c) 7-component ($R^2 = 0.68$, NMAE = 2.87\%) alloys, obtained by augmenting the $\mathcal{D}_4$-trained base model with 20 actively selected samples from the respective target domains. Error bars denote GPR predictive uncertainty ($\pm 1\sigma$). The systematic improvement with increasing chemical complexity is consistent with the unseen-element fraction analysis in Supplementary Figure~\ref{fig:unseen_fraction}.}
\end{figure*}

\subsection*{Generalization to Higher-Order Systems}
To rigorously test the physical fidelity of the pseudo-density descriptor, we evaluated the model's ability to extrapolate to the broader chemical space $\mathcal{D}_7$. We first established a robust base model trained on a budget of 200 samples from the 4-component domain $\mathcal{D}_4$. While the previous section demonstrated that 10 samples suffice for in-domain screening, this larger training budget was employed here to ensure the model fully captures the electronic manifold's topology prior to extrapolation. This distinct step allows us to decouple the quality of the descriptor from data scarcity effects during zero-shot testing. As shown in Supplementary Figure \ref{fig:supp3}, this base model exhibits non-trivial zero-shot predictive power on the 5-, 6-, and 7-component alloys of $\mathcal{D}_7$ (Mo-Nb-Ta-Ti-V-W-Zr), despite the presence of four elements (Mo, Ta, V, W) never encountered during training. This confirms the existence of a transferable electronic manifold.

However, to achieve high-fidelity quantitative predictions suitable for materials screening, we employed a few-shot domain adaptation strategy. The base model was augmented with a minimal batch of 20 actively selected samples from the respective higher-order domains. As summarized in Table \ref{tab:extrapolation} and illustrated in Figure \ref{fig:parity567}, this rapid adaptation yields significant predictive accuracy. For the 5-component alloys, the model achieves an $R^2$ of 0.85 and a NMAE of 5.54\%. As the chemical complexity increases to 7 components, the model maintains a strong correlation ($R^2 = 0.68$) and a low NMAE of 2.87\%. Visually confirmed in Figure \ref{fig:extrapolation_parity}, the predictions track the $y=x$ line closely. The accompanying uncertainty distribution (Figure \ref{fig:extrapolation_parity}, right panel) shows a peaked distribution of predicted standard deviations. Quantitative calibration on the $\mathcal{D}_7$ test set yields coverage of 43.2\% within $\pm 1\sigma$ and 78.8\% within $\pm 2\sigma$, indicating moderate overconfidence in the extrapolative regime — an expected consequence of predicting outside the training chemical space where the GP posterior uncertainty is governed by prior assumptions rather than observed data. Per-complexity calibration improves systematically with chemical complexity: 40.9\%/76.3\% (5-component), 51.6\%/88.0\% (6-component), and 67.3\%/100.0\% (7-component) within $\pm 1\sigma$/$\pm 2\sigma$ respectively, consistent with the unseen-element fraction analysis in Supplementary Figure~\ref{fig:unseen_fraction}.

\begin{figure*}[h]
    \centering
    \includegraphics[width=0.8\textwidth]{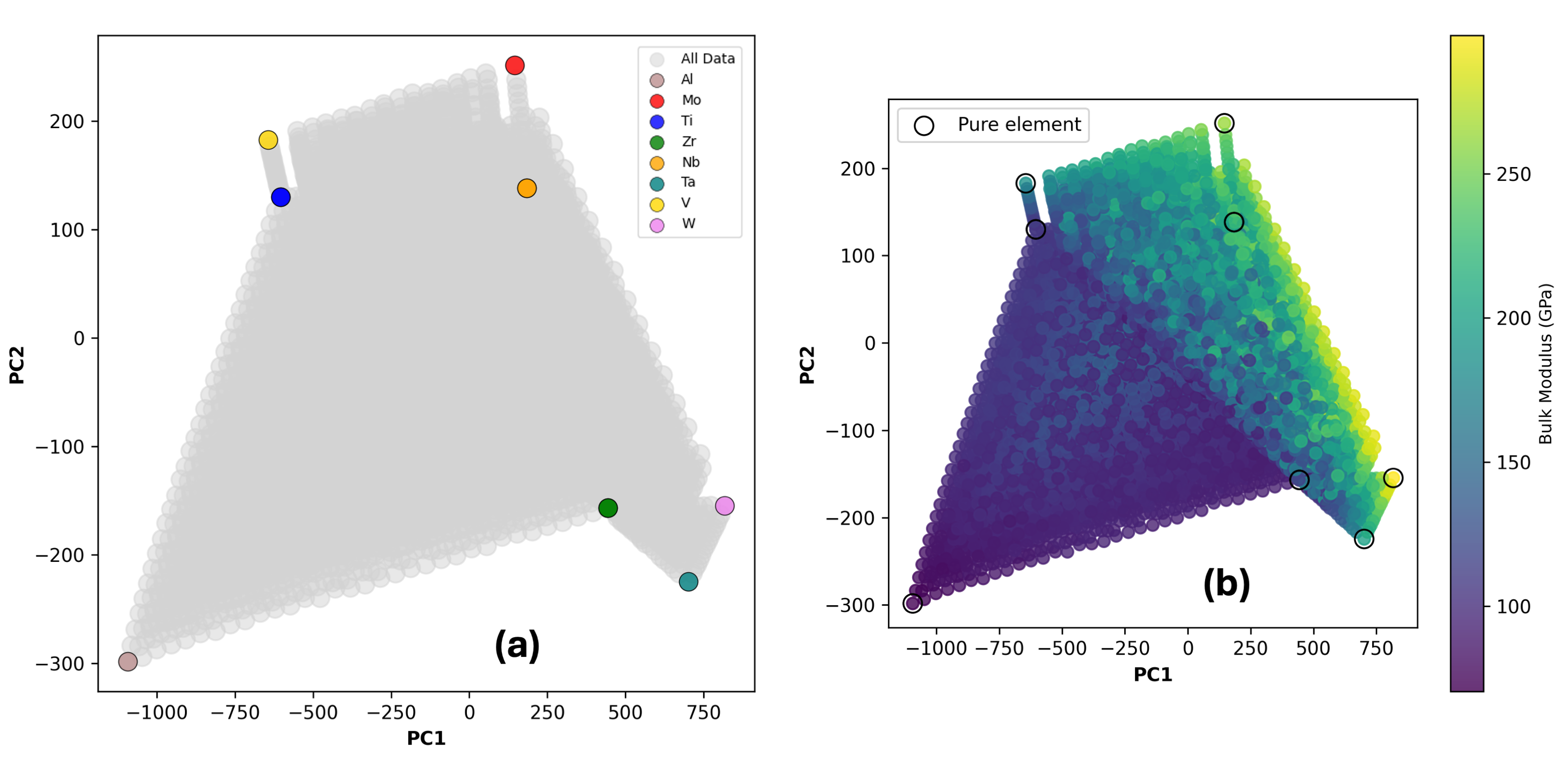}
    \caption{\textbf{Principal Component analysis of the combined $\mathcal{D}_4 + \mathcal{D}_7$ compositional space.} (a) 2D projection of pseudo-density feature vectors for the training domain ($\mathcal{D}_4$) and extrapolation domain ($\mathcal{D}_7$), with pure elemental compositions indicated. (b) The same projection colored by bulk modulus. Note that both datasets are projected onto a common PCA basis estimated jointly from $\mathcal{D}_4 + \mathcal{D}_7$; the observed overlap therefore reflects partly the shared basis construction. The physically meaningful interpretation, that the pseudo-density feature scales of the two alloy families are compatible, is confirmed independently by the $\mathcal{D}_4$-only PCA in Figure~\ref{fig:PC4only}.}
    \label{fig:supp1}
\end{figure*}

This result directly addresses a central limitation of many descriptor-based machine learning approaches in materials science, namely the difficulty of extrapolating beyond the chemical species present in the training set. Unlike composition-based descriptors that require explicit knowledge of all constituent elements, the pseudo-density descriptor operates on a transferable electronic packing manifold. Because the local valence overlap environments in $\mathcal{D}_7$ structurally resemble those in $\mathcal{D}_4$ (confirmed by the PC-space overlap in Supp. Fig. \ref{fig:supp1}), the regression mapping adapts rapidly to the new chemical labels with minimal data. 

To provide a more granular assessment of the framework's extrapolative capability, Supplementary Figure~\ref{fig:unseen_fraction} reports NMAE as a function of the unseen-element mole fraction $f_\text{unseen} = x_\text{Mo} + x_\text{Ta} + x_\text{V} + x_\text{W}$, binned by quartile across $\mathcal{D}_7$. The result exhibits a monotonically decreasing NMAE from 10.17\% at low $f_\text{unseen}$ (Q1: compositions dominated by the shared elements Nb, Ti, Zr) to 3.86\% at high $f_\text{unseen}$ (Q4: compositions predominantly constituted by the four novel elements). This trend is physically interpretable: compositions with low unseen-element fractions occupy a mixed region of the PC manifold where the $\mathcal{D}_4$-to-$\mathcal{D}_7$ transition in feature space is most abrupt, whereas compositions with high Mo+Ta+V+W content lie in a region whose d-electron autocorrelation signature is coherently distinct and well-separated, permitting more reliable regression. The superior accuracy in the high-$f_\text{unseen}$ regime is further confirmed by the parity plot for Mo/Ta/V/W-dominated compositions in Supplementary Figure~\ref{fig:motavw_parity}, which demonstrates that the pseudo-density descriptor generalizes to the fully novel compositional regime without systematic bias.

\begin{table}[h]
\small
\centering
\caption{\label{tab:extrapolation} Extrapolative performance metrics. The model was trained \textit{only} on 4-component Al-Nb-Ti-Zr and tested on 5, 6, and 7-component alloys containing Mo, Ta, V, and W.}
\begin{tabular}{lcccc}
\toprule
Components & N (Test Samples) & NMAE (\%) & MAE (GPa) \\
\midrule
5 & 3675 & 5.54 & 9.20 \\
6 & 882 & 4.16 & 7.01 \\
7 & 49 & 2.87 & 4.89 \\
\bottomrule
\end{tabular}
\end{table}

\section*{Conclusions}

We demonstrated that the non-interacting electron density offers a rigorous and scalable alternative to standard DFT-based descriptors for HEA discovery. By decoupling descriptor generation from the SCF bottleneck, this framework effectively reduces the computational investment of high-throughput screening. The integration of this descriptor with Bayesian active learning demonstrated superior sample efficiency in the Al-Nb-Ti-Zr RHEA system. The framework achieved a NMAE of <2\% for the bulk modulus using only 10 actively selected training samples, surpassing the efficiency of state-of-the-art benchmarks that rely on fully converged densities. Beyond sample efficiency within the training domain, we demonstrated the model's rigorous capability for chemical extrapolation. A model trained exclusively on the quaternary Al-Nb-Ti-Zr system successfully predicted the bulk moduli of a distinct 7-component system (Mo-Nb-Ta-Ti-V-W-Zr) containing four elements entirely absent from the training set. Furthermore, we demonstrated that augmenting this base model with just 20 samples from the target domain recovers high predictive fidelity (NMAE=2.87\% for 7-component alloys). This suggests that the pseudo-density descriptor captures a transferable electronic packing manifold within the refractory BCC alloy family, enabling efficient few-shot domain adaptation without element-specific feature engineering. Furthermore, we demonstrate the property transferability of this singular feature set; without modification, the descriptors accurately predict alloy formation energies (achieving an $R^2$ of 0.99 for the full compositional space for $\mathcal{D}_4$). The robustness of the approach was further validated through uncertainty quantification: bulk modulus predictions on $\mathcal{D}_4$ achieve 65.1\% and 92.3\% coverage within $\pm 1\sigma$ and $\pm 2\sigma$ respectively, consistent with a well-calibrated GPR. In the extrapolative $\mathcal{D}_7$ regime, coverage of 43.2\%/$\pm 1\sigma$ and 78.8\%/$\pm 2\sigma$ reflects expected mild overconfidence outside the training chemical space (Figure~\ref{fig:extrapolation_parity}b). For the alloy formation energy, the coverage was even closer to nominal, with 88.8\% and 99.8\% of the data falling within $\pm 1\sigma$ and $\pm 2\sigma$, respectively. Collectively, these results validate the voxelized atomic structure approach using pseudo-densities as a practical route to decoupling descriptor generation from the most computationally demanding stages of DFT.

The overall reduction in computational investment by lowering the requisite DFT calculations via active learning, facilitates the extensive exploration of vast compositional landscapes. Avenues for future research include: (i) validating transferability beyond BCC crystal structures, e.g., FCC, HCP systems; (ii) testing robustness of the pseudo-density descriptor to different SQS realizations per composition, supercell sizes, and pair-correlation matching quality, to determine whether the model learns transferable alloy physics or a regularized fingerprint of a specific disorder proxy; (iii) incorporating temperature-dependent properties via on-the-fly machine-learned molecular dynamics \cite{kumar_kohnsham_2023, timmerman_overcoming_2024}, and (iv) extending the framework to predict functional properties, e.g., thermal conductivity, oxidation resistance, critical for high-temperature applications. Nonetheless, the demonstrated capacity for zero-shot transfer within the refractory BCC alloy class, and few-shot domain adaptation with as few as 20 target-domain labels, marks a meaningful step toward sample-efficient computational screening of complex refractory alloy spaces. The framework is explicitly scoped to bulk modulus and formation enthalpy prediction for single-phase disordered BCC solid solutions; extension to FCC/HCP systems, configuration-sensitive properties, and magnetically ordered alloys remains an open direction for future validation.

\section*{Author contributions}

\textbf{Pranoy Ray}: Conceptualization, Methodology, Software, Validation, Formal analysis, Investigation, Data curation, Writing – original draft, Writing – review \& editing, Visualization. \textbf{Sayan Bhowmik}: Investigation, Software, Validation, Formal analysis, Writing – review \& editing. \textbf{Phanish Suryanarayana}: Conceptualization, Methodology, Validation, Resources, Writing – review \& editing, Supervision, Project administration, Funding acquisition. \textbf{Surya R. Kalidindi}: Conceptualization, Methodology, Validation, Resources, Writing – review \& editing, Supervision, Project administration, Funding acquisition. \textbf{Andrew J. Medford}: Conceptualization, Methodology, Validation, Resources, Writing – review \& editing, Supervision, Project administration, Funding acquisition. 

\section*{Acknowledgements}

PR and SK acknowledge support from NSF DMREF Award 2119640. SB, PS, and AJM gratefully acknowledge the support of the U.S. Department of Energy, Office of Science under grant DE-SC0023445. The authors acknowledge Dr. Matthew C. Barry for providing the dataset for this work through his PhD thesis \cite{barry_voxelized_2023-1}.

\section*{Data availability}

All scripts for feature engineering (computing pseudo electron densities \& spatial correlations), model training, and post-processing (uncertainty analysis) are available at \href{https://github.com/pranoy-ray/AlloyDiscovery}{https://github.com/pranoy-ray/AlloyDiscovery}. 

\section*{Conflicts of interest}
There are no conflicts to declare.



\balance


\newpage
\onecolumn
\section*{Supplementary Information}

\beginsupplement 

\begin{figure*}[h]
\centering
\includegraphics[width=0.70\textwidth]{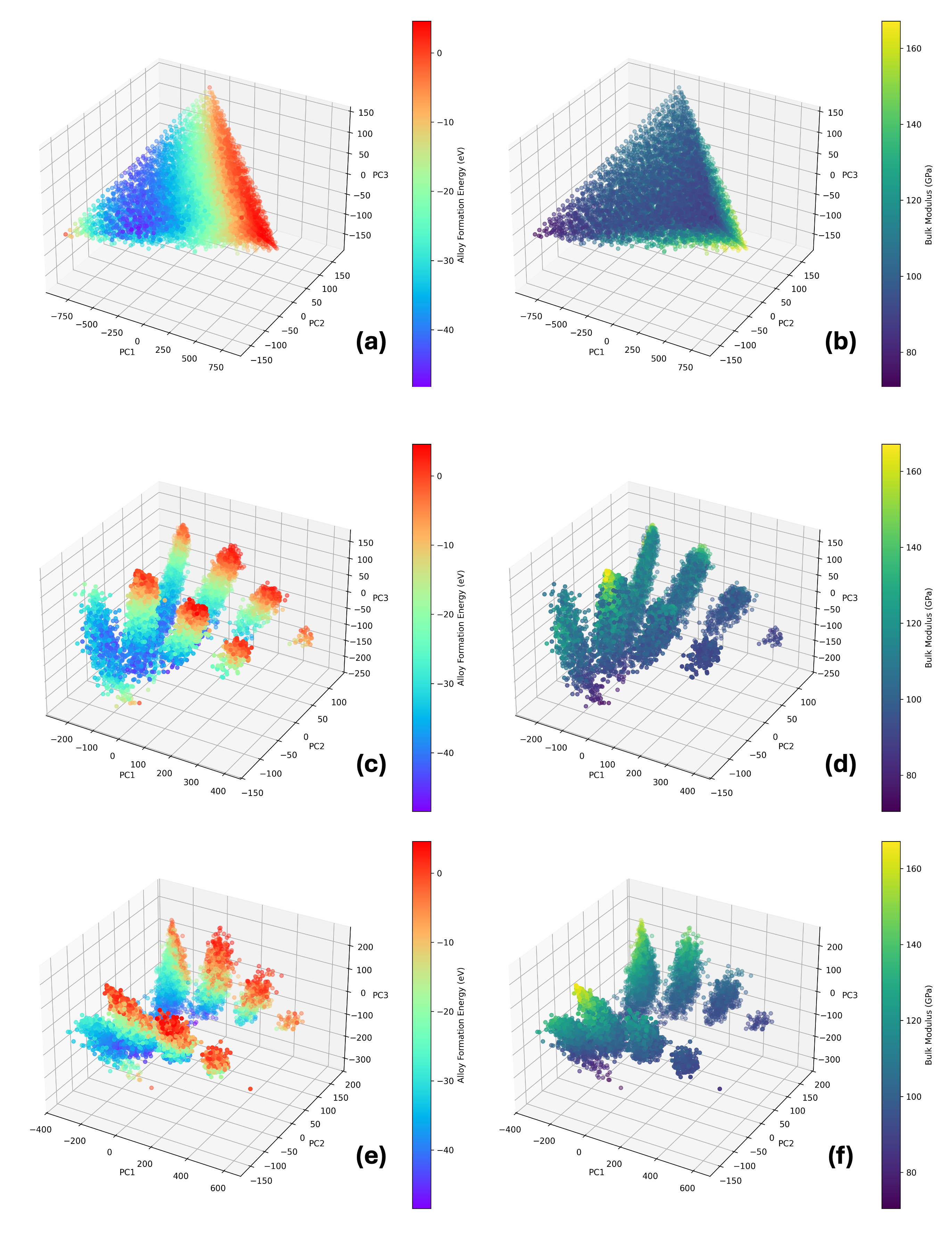} 
\caption{\label{fig:manifold_topology} Principal Component scores (colored by the respective properties) for the relaxed (a, b) SQS (unrelaxed), (c, d) VASP (relaxed), and (e, f) MACE (relaxed) structures of $\mathcal{D}_4$. The unrelaxed SQS data forms a continuous, dense simplex, whereas the relaxed structures fracture into disjoint, and distinct element-specific arms due to strain-induced geometric distortions. The smooth topology of the unrelaxed manifold acts as an engineered canonical reference state that isolates composition-driven variance; this comes at the cost of suppressing physically meaningful local distortion information, a trade-off that is acceptable for macroscopic composition-averaged properties such as bulk modulus and formation enthalpy but would not be appropriate for configuration-sensitive observables.}
\end{figure*}

\begin{figure*}[h]
    \centering
    \includegraphics[width=0.95\textwidth]{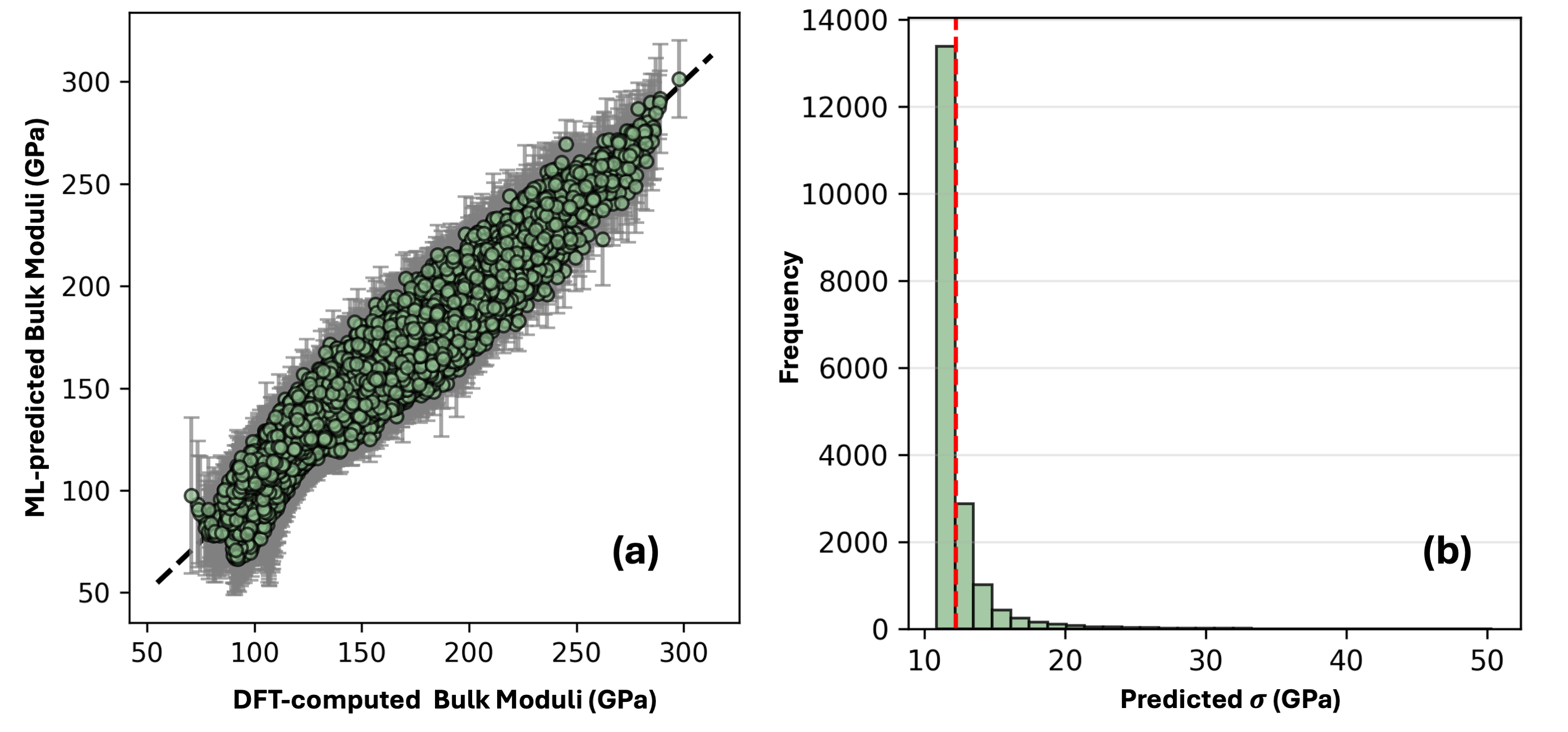}
    \caption{Uncertainty quantification for $\mathcal{D}_7$ extrapolation (random selection baseline). (a) Parity plot with $\pm 1\sigma$ error bars for bulk modulus predictions on the full $\mathcal{D}_7$ test set. (b) Distribution of predicted standard deviations $\sigma$ (GPa). Coverage of 43.2\% within $\pm 1\sigma$ and 78.8\% within $\pm 2\sigma$ indicates moderate overconfidence expected in the extrapolative regime, where GP posterior uncertainty is prior-dominated outside the training chemical space.}
    \label{fig:supp2}
\end{figure*}

\begin{figure*}[h]
    \centering
    \includegraphics[width=0.95\textwidth]{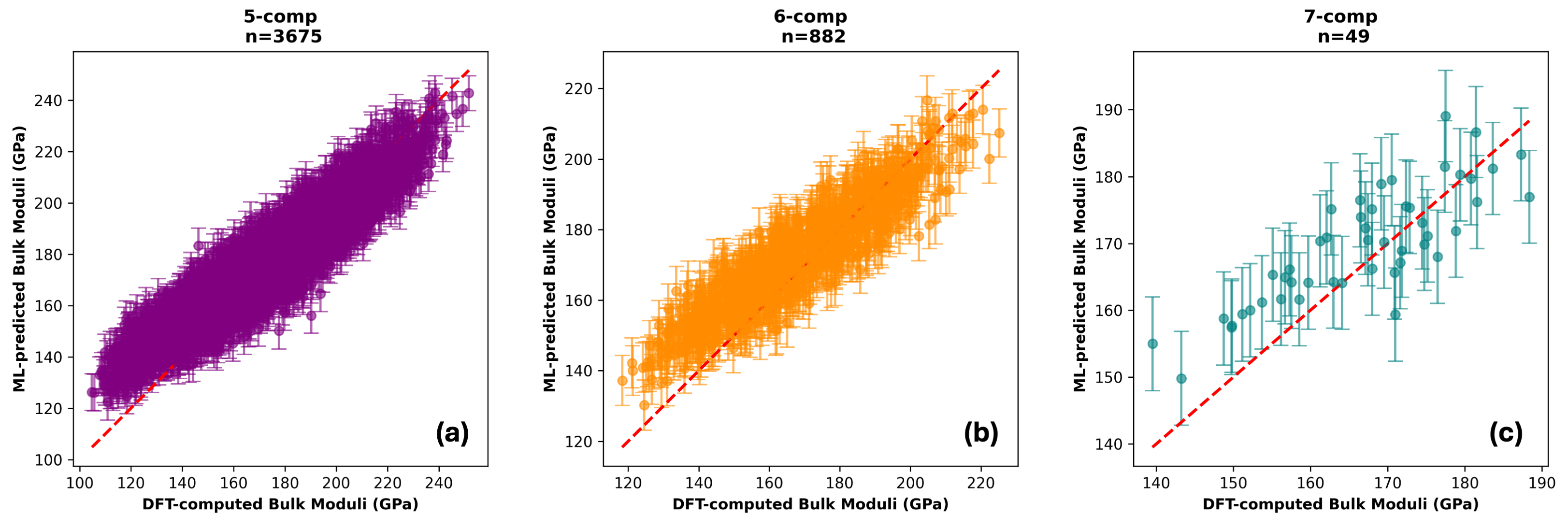}
    \caption{\textbf{Zero-shot extrapolative performance.} Predictions for 5, 6, and 7-component alloys using the base model actively trained exclusively on a budget of 200 samples from the quaternary $\mathcal{D}_4$ system, with \textit{no} exposure to Mo, Ta, V, or W. The preservation of the correlation trend ($y=x$) confirms that the physics of the refractory BCC manifold is learned solely from the lower-order system. These zero-shot predictions serve as the prior for the few-shot adaptation shown in the main text (Figure \ref{fig:parity567}).}
    \label{fig:supp3}
\end{figure*}

\begin{figure*}[h]
    \centering
    \includegraphics[width=0.75\textwidth]{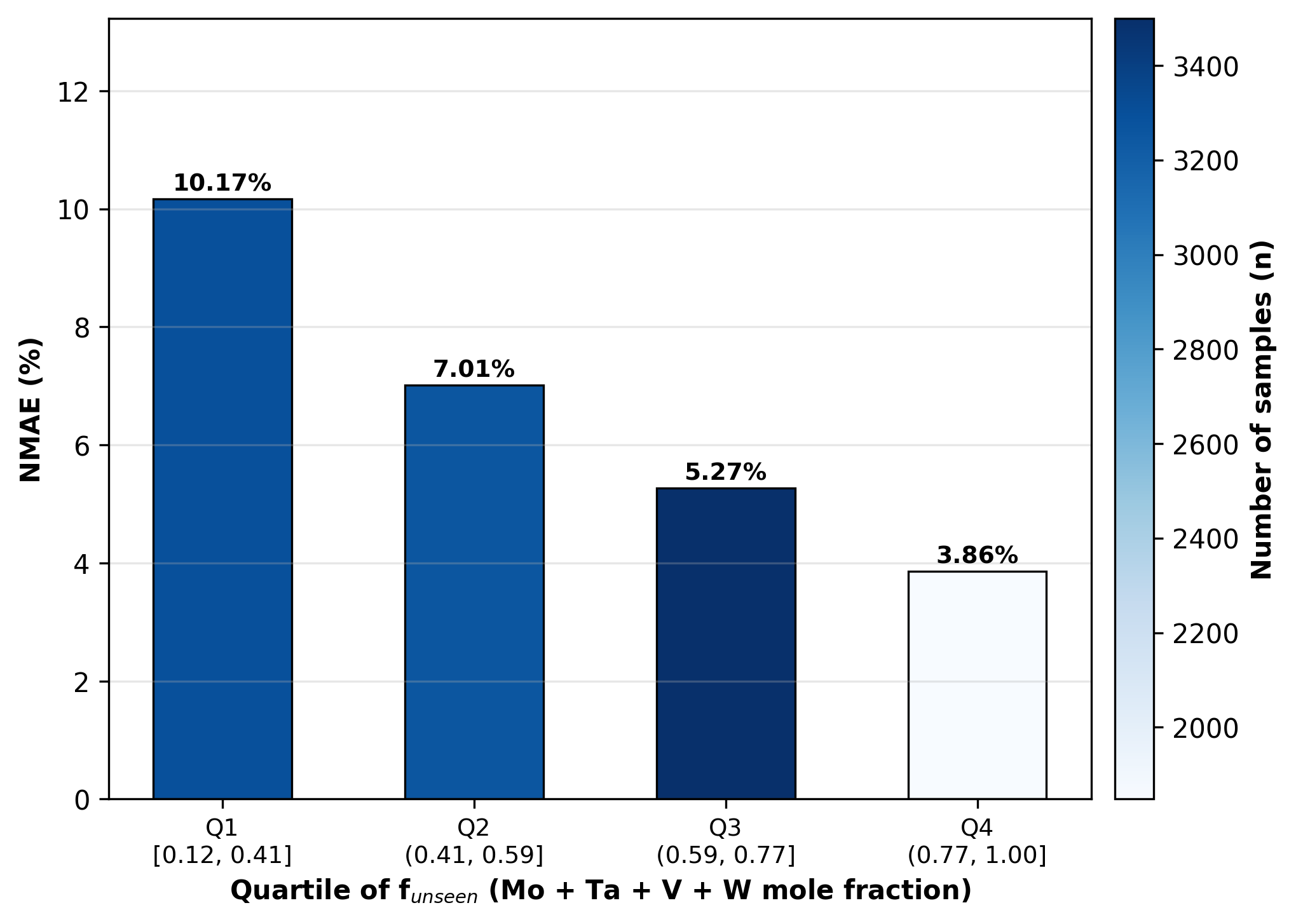}
    \caption{Extrapolative accuracy as a function of unseen-element content. NMAE for bulk modulus predictions on $\mathcal{D}_7$ compositions binned by quartile of the unseen-element mole fraction $f_\text{unseen} = x_\text{Mo} + x_\text{Ta} + x_\text{V} + x_\text{W}$. Bar color intensity encodes the number of samples per bin ($n$). NMAE decreases monotonically from 10.17\% in the lowest quartile (Q1: $f_\text{unseen} \in [0.12, 0.41]$, compositions dominated by familiar elements Nb, Ti, Zr) to 3.86\% in the highest quartile (Q4: $f_\text{unseen} \in [0.77, 1.00]$, compositions dominated by the four elements absent from training). This monotonically decreasing trend, where accuracy is highest in the fully novel compositional regime, confirms that the pseudo-density descriptor encodes transferable valence-overlap features rather than memorizing element-specific patterns; the lower accuracy at intermediate $f_\text{unseen}$ reflects the compositionally mixed regime where neither familiar nor novel elemental signatures dominate.}
    \label{fig:unseen_fraction}
\end{figure*}

\begin{figure*}[h]
    \centering
    \includegraphics[width=0.6\textwidth]{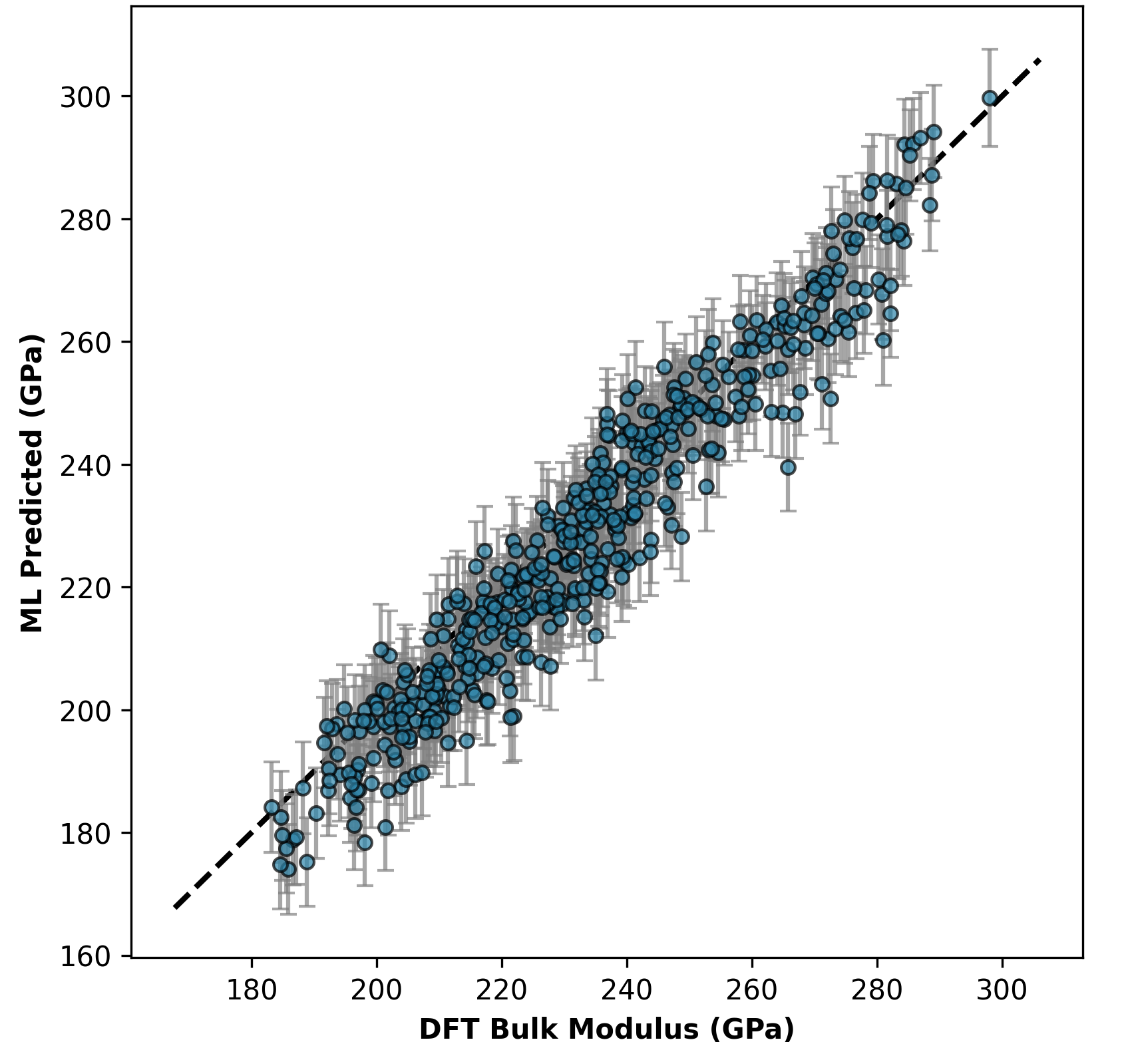}
    \caption{Parity plot for $\mathcal{D}_7$ compositions containing only unseen elements (Mo, Ta, V, W) alongside the shared elements Nb, Ti, Zr. These compositions represent the most stringent test of the framework, as their bulk modulus predictions rely entirely on the model's ability to generalize the learned valence-overlap manifold to elements with no direct training representation. The preservation of correlation with DFT ground truth confirms that the pseudo-density spatial autocorrelation encodes transferable d-electron overlap features across the group 4--6 refractory metal family.}
    \label{fig:motavw_parity}
\end{figure*}

\FloatBarrier
\begin{multicols}{2}
    \renewcommand\refname{References}
    \bibliography{rsc_dd} 
    \bibliographystyle{rsc} 
\end{multicols}

\FloatBarrier 

\end{document}